\documentclass{aa}
\usepackage[varg]{txfonts}
\usepackage[T1]{fontenc}
\usepackage{graphicx}	% Including figure files
\usepackage{xcolor}	% Including figure files
\usepackage{amsmath}	% Advanced maths commands
\usepackage{amssymb}	% Extra maths symbols
\usepackage{caption}    % for managing figures better
\usepackage{subcaption} % for managing figures better
\usepackage{multicol}  
\usepackage{booktabs}  
\usepackage{longtable}
\usepackage{footmisc}
\usepackage[breaklinks=true,colorlinks=true,linkcolor=blue,citecolor=blue,urlcolor=blue]{hyperref}
\usepackage{upgreek}
\usepackage{orcidlink}
\usepackage{placeins}
\bibpunct{(}{)}{;}{a}{}{,} % to follow the A&A style
\usepackage{soul} 
\usepackage{ulem} 
\usepackage{dblfloatfix}
\usepackage{float}
\usepackage{wrapfig}
\usepackage{multirow}
\usepackage{caption}
\usepackage{adjustbox}
\usepackage{enumitem}

\newcommand\blfootnote[1]{%
  \begingroup
  \renewcommand\thefootnote{}\footnote{#1}%
  \addtocounter{footnote}{-1}%
  \endgroup
}

\begin{document}

   \title{Empirical constraints on cosmic-ray scattering in NGC 1333 from dust polarisation}

    \author{
      Szu-Ting Chen\orcidlink{0009-0009-1088-9062}\inst{1,2,3}
      \and
      Ellis R. Owen\orcidlink{0000-0003-1052-6439}\inst{4,5 \star}
      \and
      Alison M. W. Mitchell\orcidlink{0000-0003-3631-5648}\inst{2}
      \and
      Shih-Ping Lai\orcidlink{0000-0001-5522-486X}\inst{1,6}
      \and
      Sheng-Jun Lin\orcidlink{0000-0002-6868-4483}\inst{7}
    }
    
    \institute{
      Institute of Astronomy, National Tsing Hua University, Hsinchu, Taiwan
      \and
      Erlangen Centre for Astroparticle Physics (ECAP), Friedrich-Alexander-Universit\"{a}t Erlangen-N\"{u}rnberg, Nikolaus-Fiebiger-Str. 2,
      D 91058 Erlangen, Germany
      \and
      Steward Observatory, University of Arizona, 933 N. Cherry Ave., Tucson, AZ 85721, USA
      \and 
      Astrophysical Big Bang Laboratory (ABBL), RIKEN Pioneering Research Institute (PRI), Wak\={o}, Saitama 351-0198, Japan
      \and
      Theoretical Astrophysics, Department of Earth and Space Science, Graduate School of Science, The University of Osaka, Toyonaka, Osaka 560-0043, Japan
      \and
      Center for Informatics and Computation in Astronomy (CICA), National Tsing Hua University, Hsinchu, Taiwan
      \and
      Academia Sinica Institute of Astronomy and Astrophysics, No.1, Sec. 4., Roosevelt Road, Taipei 10617, Taiwan 
    }

\offprints{S.-T. Chen, \email{szu.tingting@arizona.edu}; E. R. Owen, \email{ellis.owen@riken.jp}; A. M. W. Mitchell, \email{alison.mw.mitchell@fau.de}.}
   \date{Received ??; accepted ??}
 
  \abstract
    %Context
  {Cosmic rays (CRs) regulate molecular-cloud ionization and
chemistry. Gyro-resonant scattering by magnetic fluctuations
sets their mean free path and, on larger scales, their
effective diffusion coefficient. However, coefficients
inferred from CR distributions averaged over cloud to
Galactic scales leave local magnetic structure unresolved.} 
    {We use the magnetic-field structure of NGC\,1333 and its
embedded protostellar core IRAS\,4A to estimate local CR
scattering rates and effective diffusion coefficients,
$D(E)$, and examine their variation with environment
and scale.}  
    {We analyse JCMT/POL-2 maps of nine NGC\,1333 regions and
ALMA Band~7 maps of two IRAS\,4A regions. We estimate
plane-of-sky field strengths using the
Davis--Chandrasekhar--Fermi method and derive fluctuation
spectra from polarisation-angle structure functions.
Where these spectra are sufficiently constrained,
we apply quasi-linear gyro-resonant theory to calculate
pitch-angle scattering rates and $D(E)$.} 
{The measured fluctuations constrain scattering over
proton energy intervals of approximately 0.24--5.3~PeV
in NGC\,1333 and 10--57~TeV in IRAS\,4A.
At representative energies within these intervals,
$D(E)$ spans approximately
$3\times10^{28}$--$10^{31}\,{\rm cm^2\,s^{-1}}$.
Scattering varies substantially between regions and is
stronger where the magnetic-field angular dispersion
is greater. Extrapolation to common energies suggests
scattering in IRAS\,4A is one to two orders of magnitude
stronger than in its surroundings.
Within the directly probed intervals, the scattering
mean free paths exceed the individual region sizes.
The inferred diffusion coefficients therefore characterise local
scattering conditions. A cloud-scale diffusive regime,
if established, requires CRs to sample multiple regions
with different scattering rates.} 
    {Dust polarisation observations provide local constraints
on CR scattering in magnetically structured clouds, 
    connecting local scattering to cloud-scale transport at energies accessible to $\gamma$-ray observatories.}

   \maketitle

\blfootnote{$^{\star}$Initial stage of this work was conducted while E.R.O. was at the
Center for Informatics and Computation in Astronomy, National Tsing Hua University, Hsinchu 30013, Taiwan.}
\vspace{-1.5cm}

\section{Introduction}
\label{sec:intro}

Cosmic rays (CRs) are an important source of ionization in molecular clouds. They have long been regarded as one of the few feedback agents capable of penetrating the densest regions, which are shielded from most ionizing interstellar radiation~\citep{Goldsmith1978, Goldsmith2001ApJ}. By sustaining a residual ionization fraction in dense gas, CRs influence the coupling between gas and magnetic fields~\citep[e.g.][]{Bergin1999ApJ, Padovani2011A&A, Bailey2017A&A}, regulate ion-neutral chemistry~\citep{Desch2004ApJ, Indriolo2015, Gaches2019}, and contribute to the thermal balance of star-forming material~\citep[e.g.][]{Spitzer1968ApJ}. 
Despite their importance, 
  our understanding of the microphysics regulating CR propagation through molecular clouds is incomplete. 
  
  In dense and/or weakly ionised clouds, CR transport is traditionally considered to transition from diffusion to ballistic propagation as gas density increases~\citep{Lagage1983A&A, Felice2001ApJ, Ivlev2018ApJ, Cesarsky1978A&A}. Ion-neutral damping of resonant magnetic fluctuations reduces CR scattering in these conditions~\citep{Ivlev2018ApJ}. CR penetration then depends on energy losses and local magnetic geometry, including mirroring and focusing~\citep{Desch2004ApJ, Silsbee2018ApJ}. However, diffusive behaviour may still play a role in regulating CR propagation, for example in the surface layers of dense substructure or in intermittent magnetic fields on clump scales~\citep{Yang2023NatAs, Chernyshov2024PhRvD, Chernyshov2025PhRvD}, with potentially observable signatures~\citep{Ng2026PASJ}. 

On larger scales, the microphysics of CR propagation is unresolved. It is therefore generally treated using effective models characterised by a diffusion coefficient. 
The effective diffusion coefficients inferred from observations span more than four orders of magnitude. Estimates range from  $\approx2$--$6\times10^{26}$\,cm$^2$\,s$^{-1}$ at 1\,GeV over dense molecular clumps \citep{Yang2023NatAs}, through $\approx2\times10^{28}$\,cm$^2$\,s$^{-1}$ for GeV electrons in the star-forming disc of M 51 \citep{Heesen2023A&A}, to $\gtrsim10^{31}$\,cm$^2$\,s$^{-1}$ for 10--100\,GeV protons in the Coma galaxy cluster \citep{Wiener2019MNRAS}. These estimates probe different environments, where the underlying transport microphysics need not be the same \citep[for reviews, see][]{Owen2023Galax, Ruszkowski2023A&ARv}. The diagnostics also differ. CR secondary-to-primary ratios, synchrotron and $\gamma$-ray emission, 
and ionization tracers 
probe different CR species, energies, and propagation scales \citep[e.g.][]{Strong2007ARNPS, Padovani2009A&A, Abeysekara2017Sci, Heesen2021Ap&SS, Luo2026arXiv}. It is therefore difficult to separate variations in CR propagation from the effects of gas density 
\citep{Abrahams2017ApJ, Lai2026PhRvD}, magnetic-field structure
\citep{Beaumont2013ApJ, Crutcher2012ARA&A}, astrochemistry
\citep{Latrille2025A&A, Wakelam2005A&A, Luo2024A&A}, and projection 
\citep{Beaumont2013ApJ}.  

An alternative to these effective treatments is to follow individual CR particle trajectories through prescribed realisations of turbulent magnetic fields. Such test-particle simulations resolve gyro-scale wave-particle interactions directly and provide a stringent test of analytic transport theory~\citep[e.g.][]{Giacalone1999ApJ, Casse2001PhRvD, Mertsch2020Ap&SS}. However, their dynamic range is limited, and the turbulence must be specified a priori. They are therefore not tied to the magnetic structure of a particular cloud.

This leaves a methodological gap between microphysical test-particle simulations and the effective diffusion coefficients used in large-scale models. The magnetic fluctuations in an individual cloud are not directly connected to the CR scattering rates or effective diffusion coefficients used to describe CR transport on larger scales. In large-scale models, the energy dependence of the diffusion coefficient is generally prescribed from a turbulence model, without being derived from the fluctuations actually present~\citep[e.g.][]{Strong2007ARNPS, Evoli2018PhRvD}. This choice 
is consequential because different CR transport prescriptions produce substantial differences in galaxy star-formation rates, outflows and circumgalactic gas~\citep{Hopkins2021MNRAS}, as well as in observables such as radio synchrotron emission~\citep{Ponnada2024MNRAS}. The scale on which these models apply a diffusion coefficient also matters. A diffusion coefficient is meaningful only on scales larger than the CR scattering mean free path it implies. Whether it applies on the scale of a clump or cloud complex
therefore depends on the scattering mean free path, even
though the scattering itself occurs on gyro scales.

Bridging these regimes requires measuring magnetic fluctuations in a specific region and estimating their power at CR-resonant scales. In quasi-linear descriptions of CR transport~\citep[e.g.][]{Jokipii1966ApJ, SA1993_1, Schlickeiser2002_book}, magnetic fluctuation power near the resonant scale sets the pitch-angle scattering rate, which in turn determines the parallel mean free path and diffusion coefficient~\citep{SA1993_2, Owen2021ApJ}. Polarised sub-mm dust emission provides a way to constrain magnetic-field structure in star-forming clouds. Emission from aligned grains traces the plane-of-sky field orientation~\citep{Planck2014A&A, Pattle2023ASPC} and has provided broad insight into magnetic fields in these environments~\citep[for reviews, see][]{Crutcher2012ARA&A, Li2021Galax}. Polarisation angle structure functions quantify angular  dispersion as a function of scale, from which the turbulent magnetic-field component can be inferred. Estimates of the field strength and turbulent linewidth provide the additional quantities needed to estimate the local CR scattering rate~\citep{Davis1951, CFmethod1953, Hildebrand2009ApJ, Houde2009ApJ, Owen2021ApJ}. 

Here, we use polarised sub-mm dust emission to investigate CR propagation in the nearby NGC~1333 molecular cloud complex and its substructure, in Perseus. This study provides a first demonstration of how measured magnetic-field fluctuations can be used to estimate region-specific CR scattering rates and corresponding effective diffusion coefficients in clouds and cores. Section~\ref{sec:obser} describes the observational target and data. Section~\ref{sec:analysis} details the polarisation analysis,
region selection, and CR scattering and diffusion calculations. We present our results in Section~\ref{sec:results}, discuss them in Section~\ref{sec:discussions}, and provide our conclusions in Section~\ref{sec:conclusions}. 

\section{Observations}
\label{sec:obser}

\begin{figure*}
    \centering
    \includegraphics[width=0.45\linewidth]{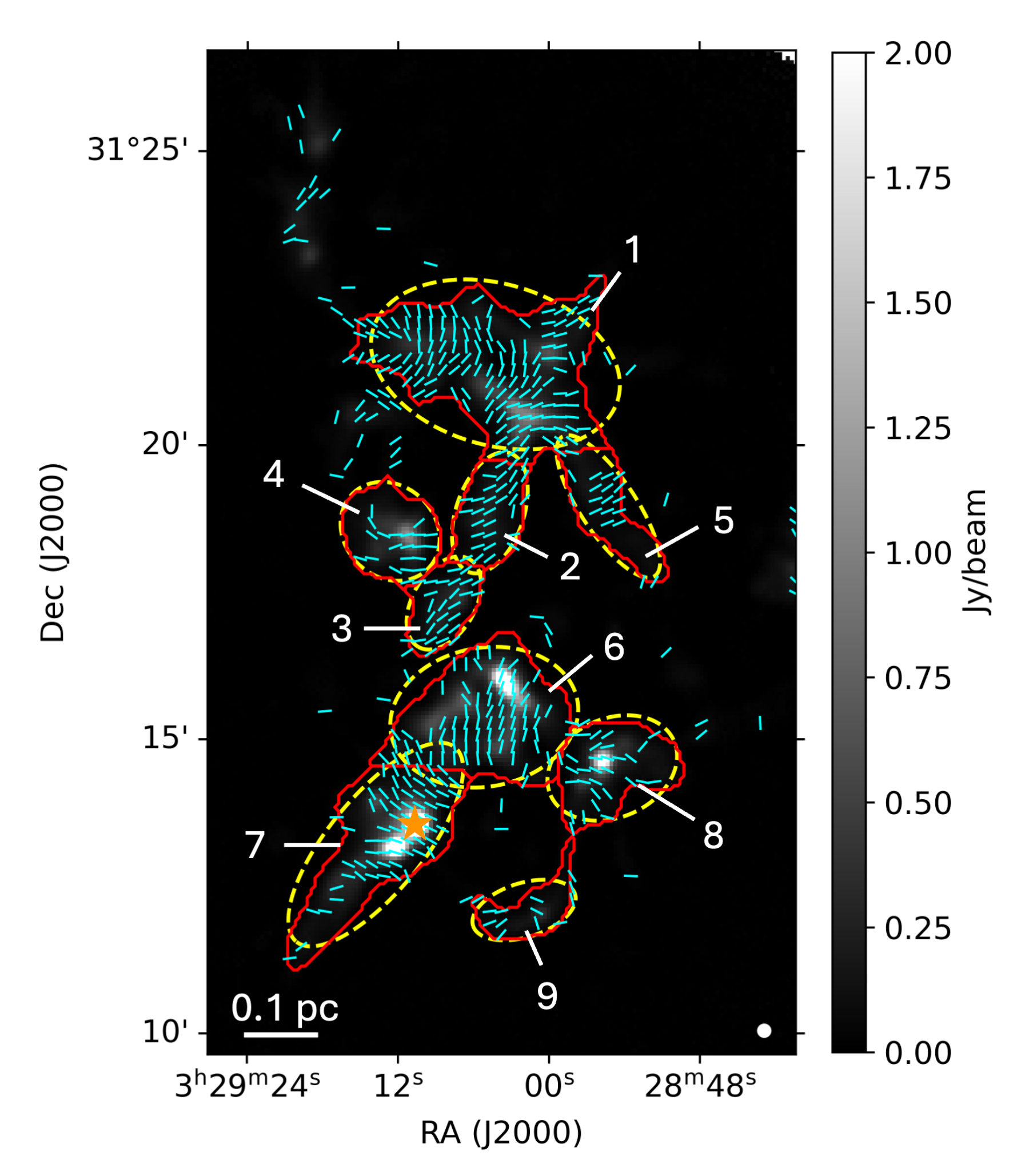}
    \includegraphics[width=0.54\linewidth]{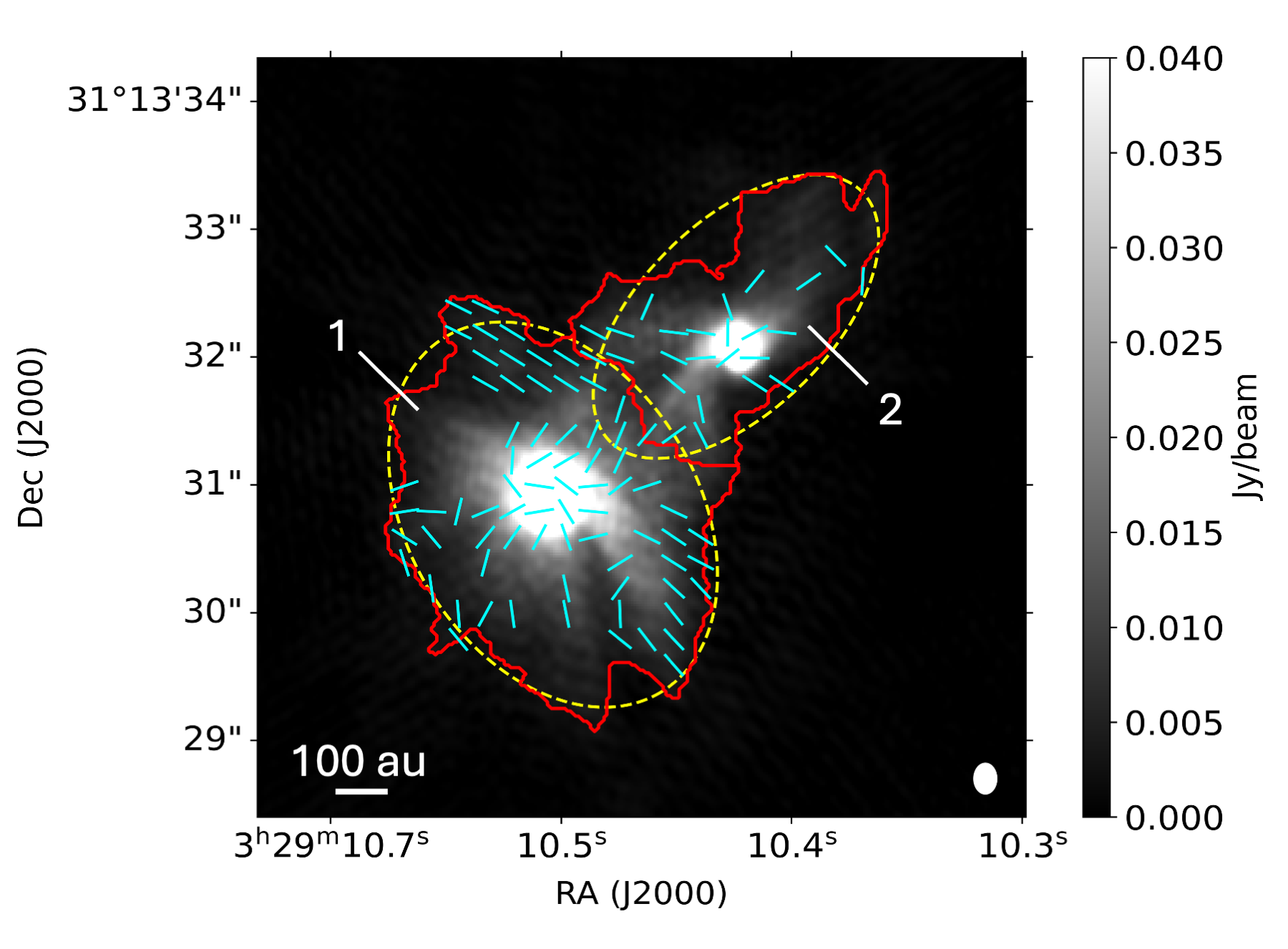}
    \caption{Stokes~I intensity maps of NGC~1333 (\textit{left}) and IRAS~4A (\textit{right}). 
    Cyan segments indicate the plane-of-sky magnetic field orientations, obtained by rotating the dust polarisation angles by 90$^\circ$. 
    Red contours represent the boundaries of the regions analyzed in this work, identified by a watershed algorithm. 
    Yellow dashed ellipses are best-fits characterizing the morphology of each region, with the fitted parameters listed in Table~\ref{tab:para1}. 
    The orange star in the NGC~1333 map marks the location of the IRAS~4A core; note that the field of view of the IRAS~4A map is much smaller than the size of this marker. 
    The scale bar and beam size of each data set are shown in the lower corner of each panel.}
    \label{fig:reg_pa}
\end{figure*}

\subsection{Target selection}
\label{sec:target_selection}

NGC\,1333 is an active region of low- and intermediate-mass star formation  
\citep{Walawender2008hsf1, Bally2008hsf1}. It hosts complex gas structure, 
proto-stellar feedback, turbulent kinematics, and a spatially varying magnetic field
\citep{Curtis2010MNRAS, Hacar2017A&A, Quillen2005ApJ, Dionatos2020A&A,
Doi2021ApJ, Chen2024MNRAS}. 
Located in the Perseus molecular cloud at a distance of $\sim294\pm 17$\,pc 
\citep{Zucker2018ApJ}, it is one of the nearest star-forming regions. 
Its proximity allows its gas and
magnetic-field structure to be studied at high spatial resolution. Additionally, recent
spatially resolved constraints on the CR ionization rate provide a
complementary probe of CR propagation through the cloud~\citep{Luo2026arXiv}. 

NGC\,1333 also hosts a rich population of young stellar objects (YSOs),
including IRAS\,4A, a young protostar in its 
earliest (Class 0) developmental stage.  High-resolution
dust-polarisation observations of IRAS\,4A are available (see
Sec.~\ref{sec:data}), allowing its magnetic-field structure to be studied in
detail. Earlier analyses have revealed a multi-scale field configuration, including an hourglass-shaped field on 100--1000\,au scales~\citep[e.g.][]{Akeson1997ApJ, Girart2006Sci, Goncalves2008A&A, Ko_2020}, and have suggested that feedback may shape the field on smaller scales. IRAS\,4A is a compact, highly resolved core embedded within the larger structure of the NGC~1333 envelope. It therefore provides a useful laboratory for localized tests of CR scattering in a different structural regime of the same cloud system. 

\subsection{Data}
\label{sec:data}

\subsubsection{JCMT data}
\label{subsec:jcmt}

We use 850~$\mu$m dust continuum observations of NGC\,1333 from the B-fields in STar-forming Region Observations (BISTRO) survey, carried out with the James Clerk Maxwell Telescope (JCMT). The polarisation measurements were obtained with the POL-2 polarimeter on the SCUBA-2 camera, which scanned the sky at 8\arcsec\,s$^{-1}$. The observations were carried out between 16 August 2017 and 16 January 2018 (Project ID: M16AL004) and were first presented by \citet{Doi_2020}. The angular resolution at 850~$\mu$m is 14\farcs1. 

Flux calibration was performed assuming a conversion factor (FCF) of 725~Jy\,pW$^{-1}$\,beam$^{-1}$. 
This was obtained by multiplying the SCUBA-2 850~$\mu$m FCF of 537~Jy\,pW$^{-1}$\,beam$^{-1}$~\citep{Dempsey_2013} by a correction factor of 1.35 for instrumental losses introduced by POL-2~\citep{Friberg_2016}. We reduced the time-series data with the \texttt{pol2map} pipeline~\citep{Currie_2014} to produce Stokes maps $I$, $Q$, and $U$ on a 4\arcsec\ pixel grid. 
The average rms noise levels in the Stokes $I$, $Q$, and $U$ maps are approximately 1.1, 0.9, and 0.9~mJy\,beam$^{-1}$, respectively. The POL-2 daisy scanning pattern\footnote{The telescope follows
a repeating pattern around the target, concentrating exposure
towards the centre of the map.} gives a non-uniform integration time and noise level across
the field. We therefore use the pixel-by-pixel uncertainties in the polarisation analysis and data selection rather than the average rms values. 

\subsubsection{ALMA data}
\label{subsec:alma}

We use archival ALMA Band~7 (345~GHz; 870~$\mu$m) polarisation observations of IRAS\,4A. The observations were obtained on 6 September 2016, with an integration time of 107 minutes (Project ID: 2015.1.00546.S). The beam size is $0\farcs25 \times 0\farcs18$. \citet{Ko_2020} performed the data calibration and imaging using the CASA software package. The rms noise levels in the Stokes $I$, $Q$, and $U$ maps used here are 0.9, 0.08, and 0.08~mJy~beam$^{-1}$, respectively. 

IRAS\,4A is a compact, bright source. The analysed regions ($\approx 1.6''$; Table~\ref{tab:para1}) lie within its inner envelope, where the emission is dominated by the compact component. Interferometric spatial filtering, commonly referred to as the missing-flux problem, increases the dispersion of polarisation angles and the polarisation fraction. This effect is strongest at low column densities and is marginal in bright cores with more ordered fields~\citep{LeGouellec2020A&A}. 
On 100--1000 au scales, the IRAS\,4A field orientations agree with those mapped at lower resolution using the Submillimeter Array~\citep{Girart2006Sci} and the Karl G. Jansky Very Large Array~\citep{Ko_2020}, which have different spatial filtering. The largest separation used in our analysis, $~\sim 1.8''$, is also below the maximum recoverable scale of these observations. These comparisons suggest that interferometric filtering is unlikely to strongly affect our results.  

\subsubsection{Molecular line data}
\label{subsec:nh3}

NH$_3$ molecular line data 
provide non-thermal linewidths used in the Davis--Chandrasekhar--Fermi (DCF;~\citealt{Davis1951, CFmethod1953}) estimates of magnetic-field strength. We use data from the Green Bank Ammonia Survey Data Release 1 (GAS DR1; \citealt{Friesen_2017}). The observations were made with the Robert C. Byrd Green Bank Telescope (GBT), with a spectral resolution of 5.7 kHz ($\sim$0.07~km~s$^{-1}$ at 23.7 GHz). The beam size is 32$\arcsec$, and the final data cube was gridded to a pixel size of $8\farcs8$. 

\begin{table*}[]
\centering
\caption{Physical parameters of the analysis regions: H$_2$ volume density, $n({\rm H_2})$, for a line-of-sight depth of 0.1 pc; non-thermal FWHM linewidth from NH$_3$, $\Delta V$; noise-corrected angular dispersion of the field orientations, $\sigma_\phi$; plane-of-sky field strength from the DCF method (Eq.~\ref{equ:DCF}), $B_{\rm pos}$; and the effective radius, $R_{\rm eff} = \sqrt{ab}$, eccentricity, $e$, and semi-major and semi-minor axes, $a$ and $b$, of the ellipse fitted to each region. The last two rows give the sizes of the region analysed with \textit{Fermi}-LAT by \citet{Jiang2025MNRAS} and of the NGC\,1333 clump in the sample of \citet{Yang2023NatAs}, used for comparison in Fig.~\ref{fig:D_vs_size}.}
\label{tab:para1}
\renewcommand{\arraystretch}{1.2}
\begin{tabular}{cccccccccc}
\hline\hline
 & Regions & n(H$_2$) & $\Delta V$ & $\sigma_\phi$ & B$_{\rm pos}$ & $R_{\rm eff}$ & $e$ & $a$ & $b$\\
 & & (cm$^{-3}$) & (km s$^{-1}$) & (deg) & ($\mu$G) & ($10^{-2}$ pc) & & ($10^{-2}$ pc) & ($10^{-2}$ pc)\\
\hline
\multicolumn{10}{c}{JCMT Observations} \\ \hline
\multirow{9}{*}{NGC\,1333}
 & 1 & 3.90$\times 10^{4}$ & 0.91 & 42.2 & 39.6 & 14.72 & 0.77 & 18.50 & 11.71\\
 & 2 & 3.47$\times 10^{4}$ & 0.57 & 21.5 & 46.2 & 6.71 & 0.84 & 9.15 & 4.92\\
 & 3 & 4.24$\times 10^{4}$ & 0.62 & 15.5 & 77.1 & 5.79 & 0.79 & 7.39 & 4.54\\
 & 4 & 2.33$\times 10^{4}$ & 0.54 & 32.1 & 24.0 & 7.00 & 0.94 & 12.15 & 4.03\\
 & 5 & 5.04$\times 10^{4}$ & 0.57 & 27.6 & 43.4 & 7.16 & 0.52 & 7.74 & 6.62\\
 & 6 & 7.75$\times 10^{4}$ & 1.00 & 20.8 & 123.8 & 11.78 & 0.68 & 13.76 & 10.09\\
 & 7 & 1.20$\times 10^{5}$ & 0.94 & 21.2 & 142.4 & 11.24 & 0.92 & 18.22 & 6.92\\
 & 8 & 3.66$\times 10^{4}$ & 0.85 & 33.8 & 44.8 & 8.38 & 0.68 & 9.79 & 7.17\\
 & 9 & 2.71$\times 10^{4}$ & 0.40 & 37.1 & 16.3 & 5.52 & 0.86 & 7.76 & 3.92\\
\hline
\multicolumn{10}{c}{ALMA Observations} \\ \hline
\multirow{2}{*}{IRAS\,4A}
 & 1 & 1.41$\times 10^{5}$ & 1.41 & 47.8 & 103.4 & 0.19 & 0.73 & 0.23 & 0.16\\
 & 2 & 1.11$\times 10^{5}$ & 1.41 & 39.6 & 110.5 & 0.14 & 0.85 & 0.20 & 0.10\\
\hline
\multicolumn{10}{c}{\textit{Fermi}-LAT Comparison (literature)} \\ \hline
\multicolumn{2}{l}{NGC\,1333 region in \citealt{Jiang2025MNRAS}} & - & - & - & - & 295 & - & 355 & 246\\
\multicolumn{2}{l}{NGC\,1333 clump in \citealt{Yang2023NatAs}} & - & - & - & - & 90 & - & 142 & 57 \\
\hline
\end{tabular}
\tablefoot{Sizes assume a distance of 294 pc~\citep{Zucker2018ApJ}. \textit{Fermi}-LAT values are taken from  \citet{Jiang2025MNRAS} and \citet{Yang2023NatAs}, and are included for comparison; they are not derived in this work.}
\end{table*}

\section{Data analysis}
\label{sec:analysis}

\subsection{Polarisation}
\label{subsec:pola}

To improve the signal-to-noise ratio (SNR) and avoid oversampling, we binned the JCMT Stokes maps to 12$\arcsec$ pixels and the ALMA maps to $0\farcs2$ pixels, close to their respective effective beam sizes. We calculated the polarisation angle ($\phi$) and fraction ($P$) from the Stokes parameters $I$, $Q$, and $U$: 
\begin{equation}
\phi = \frac{1}{2}\tan^{-1}\left(\frac{U}{Q}\right) \label{equ:Pola}
\end{equation}
\begin{equation}
P = \frac{\sqrt{Q^{2}+U^{2}-0.5\left(\sigma_{Q}^{2}+\sigma_{U}^{2}\right)}}{I},\label{equ:Pfrac}
\end{equation}
where $\sigma_Q$ and $\sigma_U$ are the rms noise levels in the Stokes Q and U maps, respectively. 

Radiative torques align the short axes of dust grains parallel to the magnetic field~\citep{Draine_1997, Lazarian2007MNRAS}. We therefore inferred the plane-of-sky magnetic-field direction by rotating the polarisation angle by $90^\circ$. We retained only measurements with $I > 3\sigma_I$ and $P > 3\sigma_P$, where $\sigma_I$ is the rms noise of the Stokes I map and $\sigma_P$ is the uncertainty in the polarisation fraction, estimated by standard error propagation. 

\subsection{Region segmentation}
\label{sec:region_segmentation}

To identify and separate structures, we applied a watershed segmentation
algorithm \citep{Beucher_Meyer_1993} to the Stokes $I$ maps. The algorithm treats each map as a topographic surface and places boundaries where basins growing from separate minima meet. This makes it suitable for separating partially connected structures with multiple peaks. 

We first smoothed the maps with a Gaussian kernel and masked them to 
$I > 3\sigma_I$. We then calculated the Euclidean distance transform of the mask, which assigns each retained pixel its distance from the nearest excluded pixel. We then applied the watershed algorithm to the negative of this transform. The negative transform has minima at the pixels furthest from the mask boundary. The watershed algorithm then grows basins from these minima and separates regions where the basins meet. This procedure separated NGC\,1333
into nine regions and IRAS\,4A into two (Fig.~\ref{fig:reg_pa};
Table~\ref{tab:para1}). 

\subsection{Empirical estimation of CR scattering and diffusion} 
\label{sec:propagation}

\begin{figure*}[h]
    \centering
    \includegraphics[width=1\textwidth]{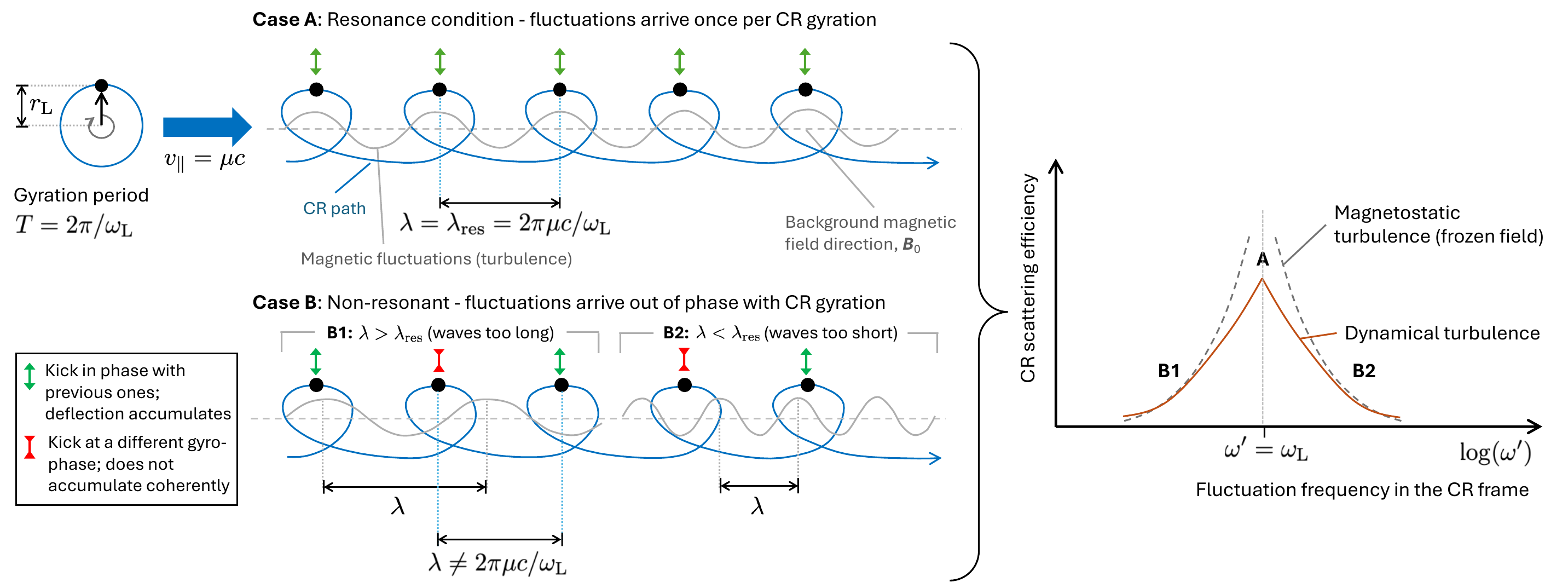} 
    \caption{Gyro-resonant interaction used to connect CR energy with
magnetic-field scale. \textit{Left:} CRs of a given energy gyrate (blue
helices; Larmor scale $r_{\rm L} = c/\omega_{\rm L}$, gyro-period
$T = 2\pi/\omega_{\rm L}$) while streaming along the background field
$\vec{B}_0$ at $v_\parallel = \mu c$ through turbulent perturbations (grey)
of wavelength $\lambda$. At the resonant wavelength, 
$\lambda_{\rm res} = 2\pi\mu c/\omega_{\rm L}$ (case A), successive
fluctuations are met at the same gyro-phase and the pitch-angle kicks add
coherently. For longer (B1) or shorter (B2) wavelengths, the phase drifts and
the kicks tend to cancel. \textit{Right:} scattering response against the
fluctuation frequency in the CR frame, $\omega' \approx |\omega -
k_\parallel \mu c|$, peaking at $\omega' = \omega_{\rm L}$, for
magnetostatic (dashed grey) and dynamical (solid orange) turbulence.} 
    \label{fig:schematic}
\end{figure*}

In molecular clouds, the high neutral gas fraction can damp wave activity, potentially reducing CR scattering and making propagation less diffusive~\citep{Lagage1983A&A, Felice2001ApJ, Draine2011piim, Ivlev2018ApJ, Dogiel2018ApJ}. Molecular cloud magnetic fields nevertheless remain structured on cloud, clump, and core scales~\citep[e.g.][]{Crutcher2012ARA&A, Li2015Natur, Li2021Galax}. Because the gyro-radii of the CRs considered here are far smaller than these structures, the particles are guided mainly along the local magnetic-field direction~\citep[e.g.][]{Schlickeiser2002_book, Yan2008ApJ, Lazarian2023FrASS}. The magnetic fluctuation power remaining at the scale resonant with each CR energy sets its pitch-angle scattering rate and, in turn, the parallel diffusion coefficient, which describes CR transport along the magnetic-field direction. We use the quasi-linear approach of \citet{SA1993_1, SA1993_2} (see also~\citealt{Owen2021ApJ}) to estimate the scattering rate from the fluctuation spectrum derived from the polarisation data, mapping each CR energy to its resonant scale.\footnote{Like other quasi-linear treatments, this approach assumes that the CR distribution is gyrotropic. Phase mixing randomises the gyrophase of the particles faster than the distribution evolves, so that it can be described by its gyrophase-averaged form, $f(\boldsymbol{x}, p, \mu, t)$ \citep{Jokipii1966ApJ, Skilling1975MNRAS_p1, Schlickeiser2002_book}.}

\subsubsection{Resonance and spatial-scale mapping}
\label{sec:res_spatial_mapping}

Gyro-resonant scattering is efficient when a CR encounters successive magnetic fluctuations at the same gyro-phase. For a given pitch-angle cosine, this selects a resonant wavelength along the local magnetic field (Fig.~\ref{fig:schematic}). 
Here,
$\mu \equiv \cos\theta$ is the pitch-angle cosine and $v \simeq c$ for
the relativistic particles considered here. We evaluate the resonance mapping at a representative pitch angle,
$\mu_\star=2/3$ ($\theta_\star\simeq48^\circ$). For an approximately
isotropic CR distribution, this is the flux-weighted mean pitch-angle
cosine of particles crossing a surface perpendicular to the local
magnetic field. This approximation assigns one effective resonant scale
to each CR energy and evaluates the pitch-angle average at a
representative value. 

We define $k \equiv |k_{\parallel}|$ as the non-negative wavenumber
parallel to the local background field, and $K_0 \equiv
|k_{\rm \parallel, res}|$ as its resonant value:
\begin{equation}
K_0(E,\mu) = \frac{\omega_{\mathrm{L}}}
{\sqrt{\mu^2 c^2 + \alpha^2 v_{\mathrm{A}}^2}} \ ,
\label{eq:k_0}
\end{equation}
\citep{SA1993_1, SA1993_2}. Here, $\alpha$ is a
dynamical-turbulence parameter. $\alpha=0$ denotes the magnetostatic
limit and $\alpha=1$ fully dynamical turbulence. For the adopted
representative pitch angle, $\mu_\star=2/3\gg v_{\rm A}/c$, so
Eq.~\ref{eq:k_0} reduces to $K_0 \simeq \omega_{\mathrm{L}}/(\mu c)$.\footnote{
\citet{Owen2021ApJ} considered the $\mu\rightarrow0$ limit, which
provides a more conservative condition for scattering. The resonant
wavelength for our adopted pitch angle is larger than in that limit by
approximately $\mu c/(\alpha v_{\rm A})$. Thus, a given range of
observed spatial scales corresponds to lower CR energies than under the
$\mu\rightarrow0$ prescription.} 

We define the resonant wavelength as
\begin{equation}
    \lambda_{\rm res}(E,\mu) \equiv \frac{2\pi}{K_0}
    \simeq \frac{2\pi\mu c}{\omega_{\rm L}}
    \simeq 2\pi\mu r_{\rm L} \ ,
\end{equation}
where $r_{\rm L}=c/\omega_{\rm L}$. For relativistic protons,
$\omega_{\rm L}\propto E^{-1}$, so $K_0$ decreases and
$\lambda_{\rm res}$ increases with CR energy. In our calculations, we
vary the CR energy to map $\lambda_{\rm res}$ across the measured
fluctuation spectrum. We adopt the low-frequency limit, where the fluctuation frequency
in the CR frame is $\omega' \simeq 2\pi\mu c/\lambda$. It follows that
$\omega'/\omega_{\rm L}\simeq\lambda_{\rm res}/\lambda$. Fluctuations
with $\lambda>\lambda_{\rm res}$ therefore lie on the low-frequency
side of the response (case B1 in Fig.~\ref{fig:schematic}), while
those with $\lambda<\lambda_{\rm res}$ lie on the high-frequency side
(case B2). Dynamical turbulence limits the interaction coherence time
and broadens the resonant response. Our calculations do not capture asymmetry between left- and
right-handed modes, so we assume zero helicity. We also assume a
locally slab-like geometry, with the background field varying on scales
larger than the resonant fluctuations, and adopt fully dynamical
turbulence ($\alpha=1$). 

\subsubsection{Estimating CR scattering from magnetic fluctuations}
\label{sec:struc_to_d_e}

We characterize the fluctuation power sampled around the
resonant scale using the statistic: 
\begin{equation}
\mathcal{J}(K_0) \equiv 
\int_0^{K_0}\frac{k}{ k_{\rm c}K_0}\,{\mathcal{P}}_\parallel(k)\,\mathrm{d}k
\;+\;\int_{K_0}^{\infty}\frac{K_0}{k_{\rm c} k}\,{\mathcal{P}}_\parallel(k)\,\mathrm{d}k \ ,
    \label{eq:j_def}
\end{equation}
where ${\mathcal{P}}_\parallel$ is 
the one-dimensional fluctuation-power spectrum as a function of the parallel wavenumber $k \equiv |k_{\parallel}|$. 
For a locally slab-like magnetic field
configuration, the CR scattering coefficient is then: 
\begin{equation}
    P_{\mu\mu}(E, \mu) \approx (1-\mu^2)\,
        \frac{\mathcal{J}(K_0) k_{\rm c}}{v_{\rm A}\lambda_{\rm td} K_0}
        \left(\frac{\omega_{\rm L} B_{\rm ref}}{B}\right)^{\!2} \mathcal{I}_\perp \ ,
    \label{eq:pmumu}
\end{equation} 
where $B$ is the local field strength estimated for each region,
$B_{\rm ref} = 1$\,mG is a reference normalisation
\citep{Crutcher2010, Li2015Natur}, and $\omega_{\rm L}$ is the CR
gyro-frequency. In $\mathcal{J}$, the resonant wavenumber is
$K_0(E,\mu)$ from Eq.~\ref{eq:k_0}; in the denominator of
Eq.~\ref{eq:pmumu}, $K_0$ takes its $\mu\to0$ value,
$\omega_{\rm L}/(\alpha v_{\rm A})$.\footnote{The leading
pitch-angle dependence, $(1-\mu^2)$, recovers the
\citet{Owen2021ApJ} form as $\mu\to0$.} 
The factor $\mathcal{I}_\perp$ accounts for magnetic-field
fluctuations perpendicular to the background field. Under the
isotropic slab approximation of \citet{Owen2021ApJ},
$\mathcal{I}_\perp = \pi/k_{\rm c}^2$, where
$k_{\rm c} = \omega_{\rm p,ref}/v_{\rm A}$ is a reference wavenumber
and $\omega_{\rm p,ref}$ is the gyro-frequency of a proton at the
reference energy $E_{\rm ref} = 100$\,MeV. This reference scale sets
the normalisation in Eqs.~\ref{eq:j_def} and~\ref{eq:pmumu}.
We set the turbulent decay length to
$\lambda_{\rm td} \approx v_{\rm A}\tau_{\rm td}$, with
$\tau_{\rm td} = 2$\,Myr~\citep{Gao2015ApJ, Larson2015ApJ}.
We take $v_{\rm A}$ to be the Alfv\'en speed of the ionized gas, 
$v_{\rm A} = B\,[4\pi\rho\,\xi/(1+\xi)]^{-1/2}$, where
$\rho = 2.8\,m_{\rm H}\,n({\rm H_2})$ is the gas density and
$\xi$ is the ionization fraction. For all regions, we adopt
$\xi = 10^{-6.5}$, the median electron fraction measured by
\citet{Pineda2024A&A} across south-eastern NGC\,1333, including
IRAS\,4A. This value lies at the upper end of the range found in dense
cores~\citep{Caselli1998ApJ}.  

We estimate the fluctuation spectrum from the binned second-order
structure function of the polarisation angles. For each region, we
calculate $S_2(\ell_j)$ in $N$ separation bins of width
$\Delta\ell$,\footnote{The bin width is $1.2\times10^{-2}$\,pc for
the NGC\,1333 regions and $2\times10^{-4}$\,pc for IRAS\,4A
(Table~\ref{tab:diffusion}).} centred on $\ell_j$. In each bin, we
subtract the mean summed measurement variance from the mean squared
difference in orientation, $S_2(\ell_j) = \langle \Delta\phi^2 \rangle_j - \langle \sigma_{\phi,a}^2 + \sigma_{\phi,b}^2 \rangle_j$, where the angle differences lie in $[0^\circ,90^\circ]$
\citep[following][]{Hildebrand2009ApJ}. For a statistically homogeneous field, the structure function and
fluctuation spectrum are related by 
\begin{equation}
S_2(\ell) = 2\int_0^\infty \mathcal{P}_\parallel(k)\,
(1-\cos k\ell)\,{\rm d}k \ . 
\end{equation}  
We estimate the spectrum by evaluating the Fourier transform
discretely: 
\begin{equation}
\mathcal{P}_\parallel(k_n) = \frac{2}{N}\left|\sum_{j=0}^{N-1} S_2(\ell_j)\,
{\rm e}^{-{\rm i}k_n\ell_j}\right| \ ,
\label{eq:Pk}
\end{equation}   
where $k_n = n\Delta k = 2\pi n/(N\Delta\ell)$ for
$n = 1,\dots,\lfloor N/2\rfloor$. Each $\mathcal{P}_\parallel(k_n)$
represents the fluctuation power integrated over the $n$th sampled
spectral bin.\footnote{We use the modulus of the transform; see
Appendix~\ref{app:limitations} for discussion.} 
With this normalisation, $\mathcal{P}_\parallel(k_n)$ is independent
of the chosen bin width. We exclude the $n=0$ term because the mean
level of $S_2$ contributes no power at $k>0$. Within the local slab
approximation, we treat the scalar Fourier modes $k_n$ inferred from
the structure function as samples of the effective parallel spectrum.  
We approximate Eq.~\ref{eq:j_def} using these discrete modes: 
\begin{equation}
\mathcal{J}(K_0) \approx k_{\rm c}^{-1}
\sum_{n=1}^{i_b}\frac{k_n}{K_0}\,{\mathcal{P}}_\parallel(k_n)
\;+\;k_{\rm c}^{-1}\sum_{n=i_b+1}^{N_{\rm modes}}\frac{K_0}{k_n}\,{\mathcal{P}}_\parallel(k_n),
    \label{eq:j_discrete}
\end{equation}
where $N_{\rm modes} = \lfloor N/2 \rfloor$ and $i_b$ is the
largest mode index for which $k_{i_b} \leq K_0$. 

The parallel spatial diffusion coefficient is generally given by 
\begin{equation}
    D_\parallel(E)=\frac{v^2}{8}
\int_{-1}^{1}
\frac{(1-\mu^2)^2}{P_{\mu\mu}(E,\mu)}
{\rm d}\mu \ .
\end{equation} 
We evaluate this integral using the representative pitch-angle cosine
defined above, approximating $P_{\mu\mu}(E,\mu)$ by
$P_{\mu\mu}(E,\mu_\star)$ with $\mu_\star=2/3$. For relativistic
particles, $v\simeq c$, and
$\int_{-1}^{1}(1-\mu^2)^2\,{\rm d}\mu=16/15$, giving 
\begin{equation}
    D(E) \approx \frac{2c^2}{15\,P_{\mu\mu}(E, \mu_\star)} \ .
    \label{eq:D}
\end{equation}
For each region, we calculate $D(E)$ from the measured
$\mathcal{P}_\parallel(k_n)$ using Eqs.~\ref{eq:pmumu},
\ref{eq:j_discrete}, and~\ref{eq:D}. We estimate its uncertainties
by bootstrap resampling the polarisation vectors and drawing the
source distance from a Monte Carlo distribution (see
Appendix~\ref{app:uncertainty}). 

\begin{figure*}[!t]
    \centering
    \includegraphics[width=0.49\linewidth]{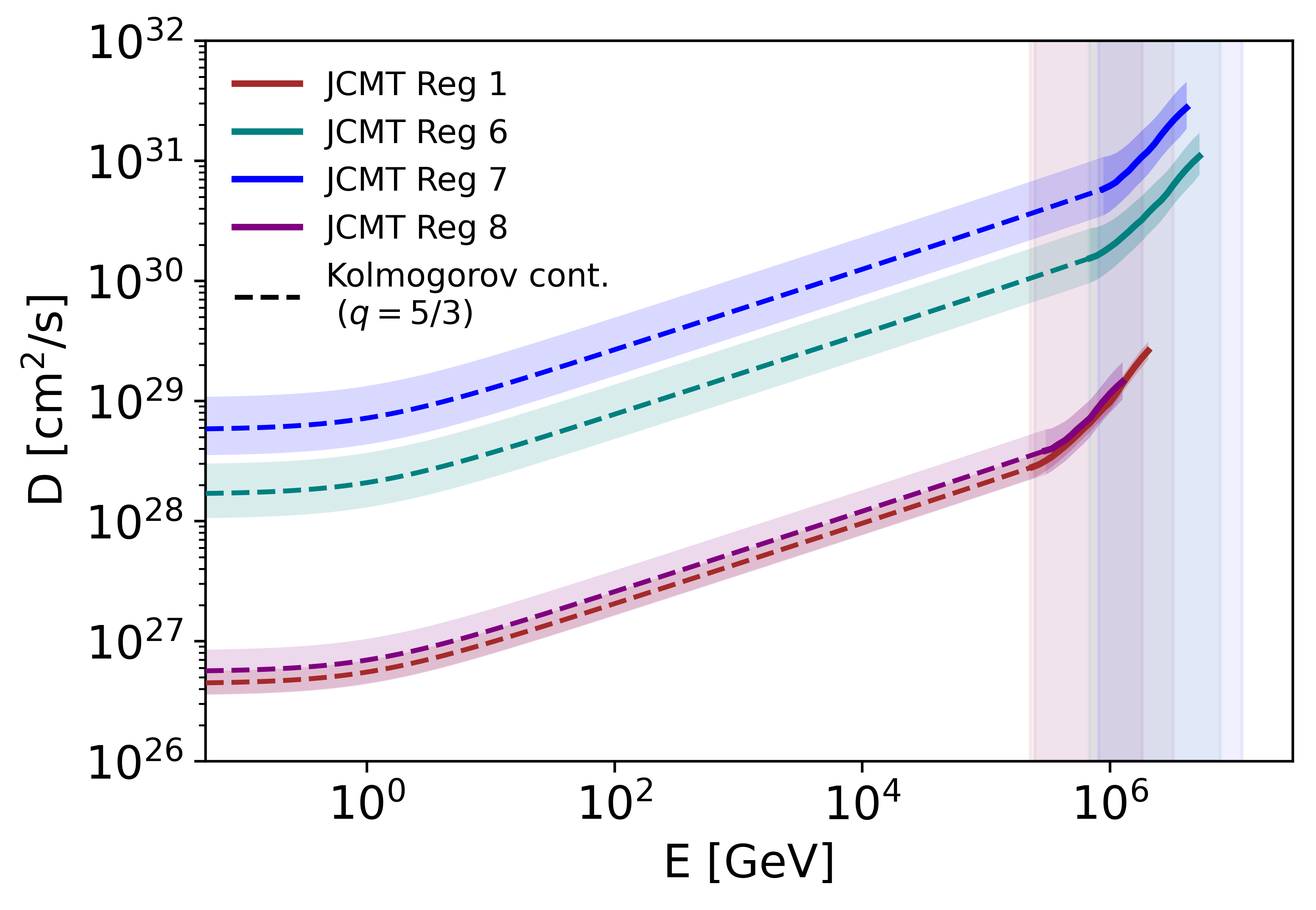}
    \includegraphics[width=0.49\linewidth]{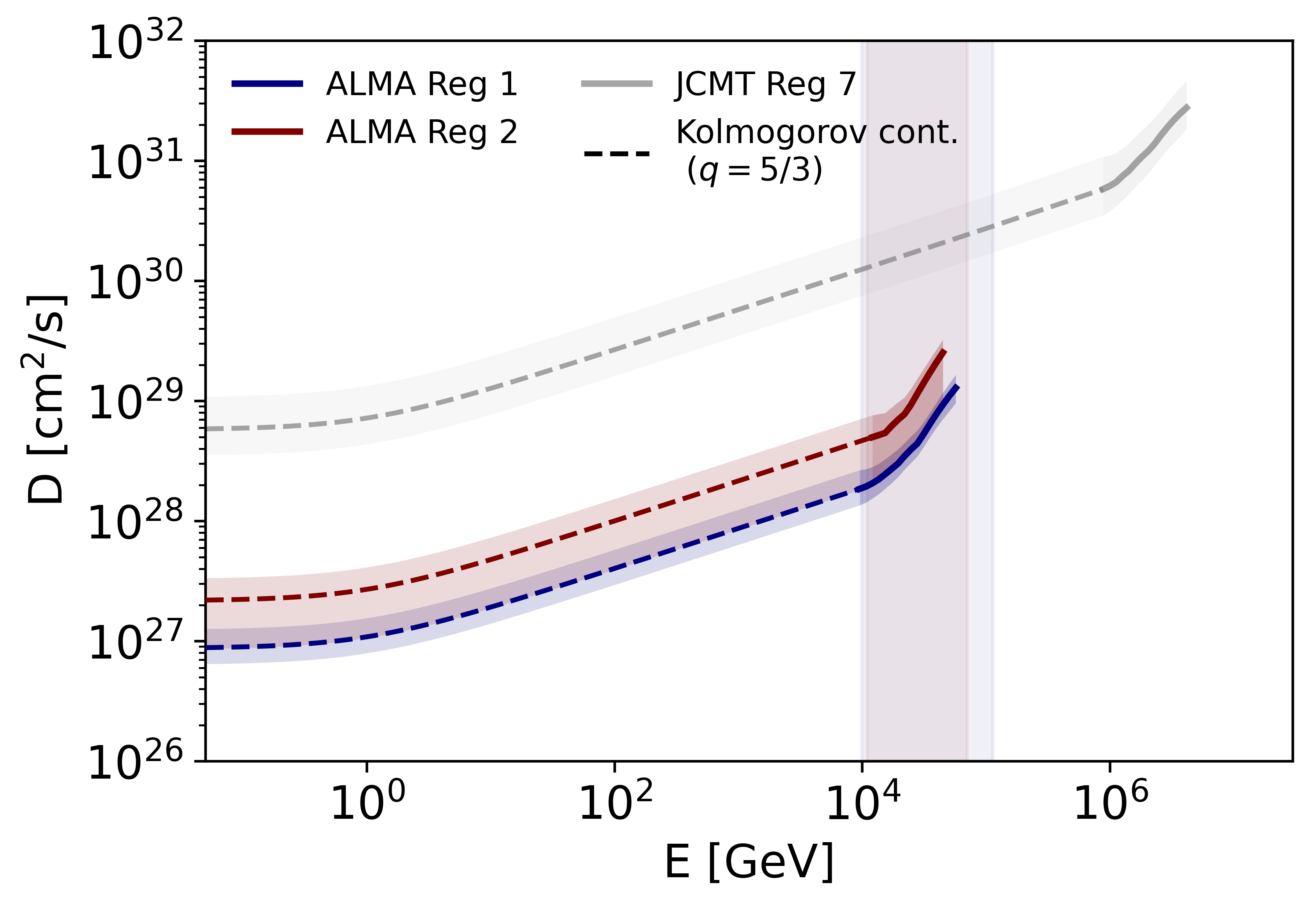} 
\caption{CR spatial diffusion coefficient, $D(E)$, for the JCMT NGC\,1333
regions (\textit{left}) and the ALMA IRAS\,4A regions
(\textit{right}). Solid segments show the data-informed intervals;
dashed segments show the Kolmogorov continuations below them.
Vertical shaded bands mark the nominal intervals, and shaded
envelopes show bootstrap uncertainties (Appendix~\ref{app:uncertainty}).
Colours identify the regions in Fig.~\ref{fig:reg_pa}. The right panel
also shows NGC\,1333 Region~7 in grey, which contains IRAS\,4A.}  
    \label{fig:D_err}
\end{figure*}

\subsubsection{Directly constrained energy range and extrapolations}
\label{sec:extrapolation_limits} 

The energy dependence of $D(E)$ enters through the CR gyro-frequency
$\omega_{\rm L}(E)$ and the corresponding resonant wavenumber
$K_0(E)$. As the energy increases, the resonant scale increases, so
different CR energies sample different parts of the measured
fluctuation spectrum. Within the sampled range, the shape of $D(E)$
is set by $\mathcal{J}(K_0)$. The polarisation data cover a finite
range of spatial scales, set by the angular resolution and the size
of each analysis region. They therefore constrain $D(E)$ directly
only over a finite energy interval.

The smallest sampled scale, $\ell_{\min}$, sets the lower-energy
boundary of this interval, while the largest sampled scale, 
$\ell_{\max}$, sets its upper-energy boundary:  
\begin{align}
E_{\rm low} & \approx 9\,{\rm TeV}
\left(\frac{B}{100\,\mu{\rm G}}\right)
\left(\frac{\ell_{\min}}{4\times10^{-4}\,{\rm pc}}\right)
\left(\frac{\mu_\star}{2/3}\right)^{-1}, \label{eq:e_band_low} \\
E_{\rm high} & \approx 8\,{\rm PeV}
\left(\frac{B}{100\,\mu{\rm G}}\right)
\left(\frac{\ell_{\max}}{0.36\,{\rm pc}}\right)
\left(\frac{\mu_\star}{2/3}\right)^{-1} \ .  
\label{eq:e_band_high}
\end{align}
Here, $\ell_{\min}$ is the effective sampling scale, taken to be
twice the separation-bin spacing (the practical Nyquist limit).
For the nominal interval, $\ell_{\max}$ is the largest separation in
the region; for the data-informed interval, it is the largest
separation retained in the analysis. These definitions give the
nominal energy interval corresponding to the full spatial range
sampled by each observation. The data-informed interval can be
narrower if some scales are excluded from the analysis. 

The relation between $\lambda_{\rm res}=2\pi/K_0$ and the sampled
spatial range gives three regimes for the finite-band calculation: 

\textbf{1. High-energy asymptote, where
$\lambda_{\rm res} > \ell_{\max}$:}
The resonant scale lies beyond the largest scale sampled by the
data. Only the second term 
in Eq.~\ref{eq:j_discrete} contributes so, from its pre-factor, 
$\mathcal{J}(K_0) \propto K_0 \propto E^{-1}$ and
$D(E) \propto E^2$. This is the asymptotic behaviour of the
finite-band calculation; it does not constrain the unresolved
fluctuation spectrum. Power on scales larger than $\ell_{\max}$,
such as cloud-scale fluctuations, would soften this high-energy
asymptote.

\textbf{2. Data-informed regime, where
$\ell_{\min} < \lambda_{\rm res} < \ell_{\max}$:}
The resonant scale lies within the sampled range. As the CR energy
increases, $K_0$ moves across the measured spectrum and
$\mathcal{J}(K_0)$ responds to the power at each scale. The measured
spectrum therefore informs the shape of $D(E)$; changes in its slope
produce corresponding bends in $D(E)$. As $\lambda_{\rm res}$
approaches $\ell_{\max}$, the contribution from larger scales (the
first term in Eq.~\ref{eq:j_discrete}) decreases and vanishes when
$\lambda_{\rm res}$ exceeds the sampled range. The curve then
approaches the high-energy asymptote, with $D(E)$ steepening towards
$E^2$.

\textbf{3. Low-energy limit, where
$\lambda_{\rm res} < \ell_{\min}$:}
The resonant scale lies below the smallest scale sampled by the data.
For a spectrum truncated at $\ell_{\min}$, the first term in
Eq.~\ref{eq:j_discrete} gives $\mathcal{J}(K_0) \propto
K_0^{-1} \propto E$, so $D(E)$ approaches a constant. None of our data-informed curves reaches this regime.
For comparisons at lower energies, we extend the fluctuation
spectrum to unresolved smaller scales using a Kolmogorov
index, $q=5/3$. For NGC\,1333, we also use the joint index
inferred from the structure functions
(Sec.~\ref{sec:field_structure}). Each continuation of $D(E)$
is anchored to the inferred value at the lower edge of its
data-informed interval.  

The full spatial ranges sampled by the ALMA data toward IRAS\,4A
($\approx 4\times10^{-4}$--$5\times10^{-3}$\,pc) and the JCMT data across
NGC\,1333 (0.02--0.36\,pc) correspond to nominal resonant proton
energy ranges of $\approx 10$--110\,TeV and $\approx 0.23$--12\,PeV,
respectively. The data-informed intervals are narrower where our  
analysis excludes some of the largest separations (Sec.~\ref{sec:results}).

\section{Results}
\label{sec:results}

\subsection{Magnetic field strength}
\label{sec:strength}

We derive the plane-of-sky magnetic field strength using the Davis-Chandrasekhar-Fermi (DCF) method \citep{Davis1951, CFmethod1953}, which is widely adopted for estimating magnetic field magnitudes: 
\begin{equation}
    B_{\rm pos} = Q\sqrt{4\pi\rho}\frac{\sigma_v}{\sigma_\phi}=9.3\sqrt{n(\rm H_2)}\frac{\Delta V_{\rm nt}}{\sigma_\phi} \ . \label{equ:DCF}
\end{equation}
Here, $Q$ is a correction factor (0.5; \citealp{Ostriker2001}), $\rho$ is the gas density, $\sigma_v$ is the turbulent velocity dispersion, $\sigma_\phi$ is the magnetic field angular dispersion, $n(\rm H_2)$ is the volume density, and $\Delta V_{\rm nt}$ is the non-thermal FWHM linewidth. For the numerical expression, $B_{\rm pos}$ is in $\mu$G,
$n({\rm H_2})$ in ${\rm cm^{-3}}$, $\Delta V_{\rm nt}$ in 
${\rm km\,s^{-1}}$, and $\sigma_\phi$ in degrees. 

The H$_2$ volume density is derived from the dust continuum assuming a
line-of-sight depth of 0.1~pc, while $\sigma_\phi$ is the
noise-corrected dispersion of the field orientations and
$\Delta V_{\rm nt}$ is obtained from the NH$_3$(1,1) linewidth.
Details of these measurements are given in
Appendix~\ref{app:field_strength}. The resulting parameters and field
strengths for each region are listed in Table~\ref{tab:para1}. 
The measured $\sigma_\phi$ also includes coherent curvature of the 
ordered field, which can lead to an underestimate of $B$ by up to a
factor of approximately two (Appendix~\ref{app:limitations}).

\subsection{Magnetic Field Structure}
\label{sec:field_structure}

The plane-of-sky magnetic-field orientations differ across the two
datasets (Fig.~\ref{fig:reg_pa}). At the larger scales probed by JCMT,
the fields are broadly ordered, although their orientations vary
between regions and, in some cases, across a region. ALMA resolves the
curved, approximately hourglass-shaped field of IRAS~4A
\citep{Girart2006Sci,Ko_2020}. The corresponding second-order
structure functions, $S_2(\ell)$, and fluctuation-power spectra,
$P_\parallel(k)$, are shown in Appendix~\ref{app:results}. The JCMT spectra have broadly similar shapes but differ in amplitude, whereas the two IRAS 4A spectra also differ in shape.  

For the JCMT regions, we retain vector pairs with separations no
greater than the full minor axis, $2b$, of the fitted ellipse
(hereafter the \textit{lag cap}). Larger separations sample only
structure along the major axis and introduce window-edge ringing in
$\mathcal{P}_\parallel(k)$ (Appendix~\ref{sec:modes}). We also require
more than three retained spectral modes, so that the data-informed
interval spans at least a factor of four in wavenumber. This
criterion was set before examining the regional results and excludes
Regions 2, 3, 4, 5, and 9 from the diffusion analysis.

For IRAS\,4A, $S_2$ decreases at the largest separations because pairs
that span the hourglass field can have similar orientations. We
therefore remove the first bin that lies more than $2\sigma$ below
the maximum at smaller separations, together with all larger
separations (hereafter the \textit{trim}). We also require the
inferred $D(E)$ to remain stable when one additional bin is removed
(Appendix~\ref{sec:modes}). These selections lower the upper edge of
the data-informed energy interval. Coherent field curvature may
still contribute to the inferred fluctuation power at the largest
retained scales (Appendix~\ref{app:limitations}). Fits to the direction-balanced structure functions give the
regional spectral indices listed in Table~\ref{tab:diffusion}.
The four retained NGC\,1333 regions are consistent with a
shared index of $q=1.38^{+0.08}_{-0.17}$
(Appendix~\ref{sec:modes}). The IRAS\,4A structure functions
are saturated at the resolved scales, so no reliable index
can be measured there. 

\subsection{CR diffusion coefficient}
\label{sec:diff_results}

\begin{figure}[!t]
    \centering
    \includegraphics[width=\columnwidth]{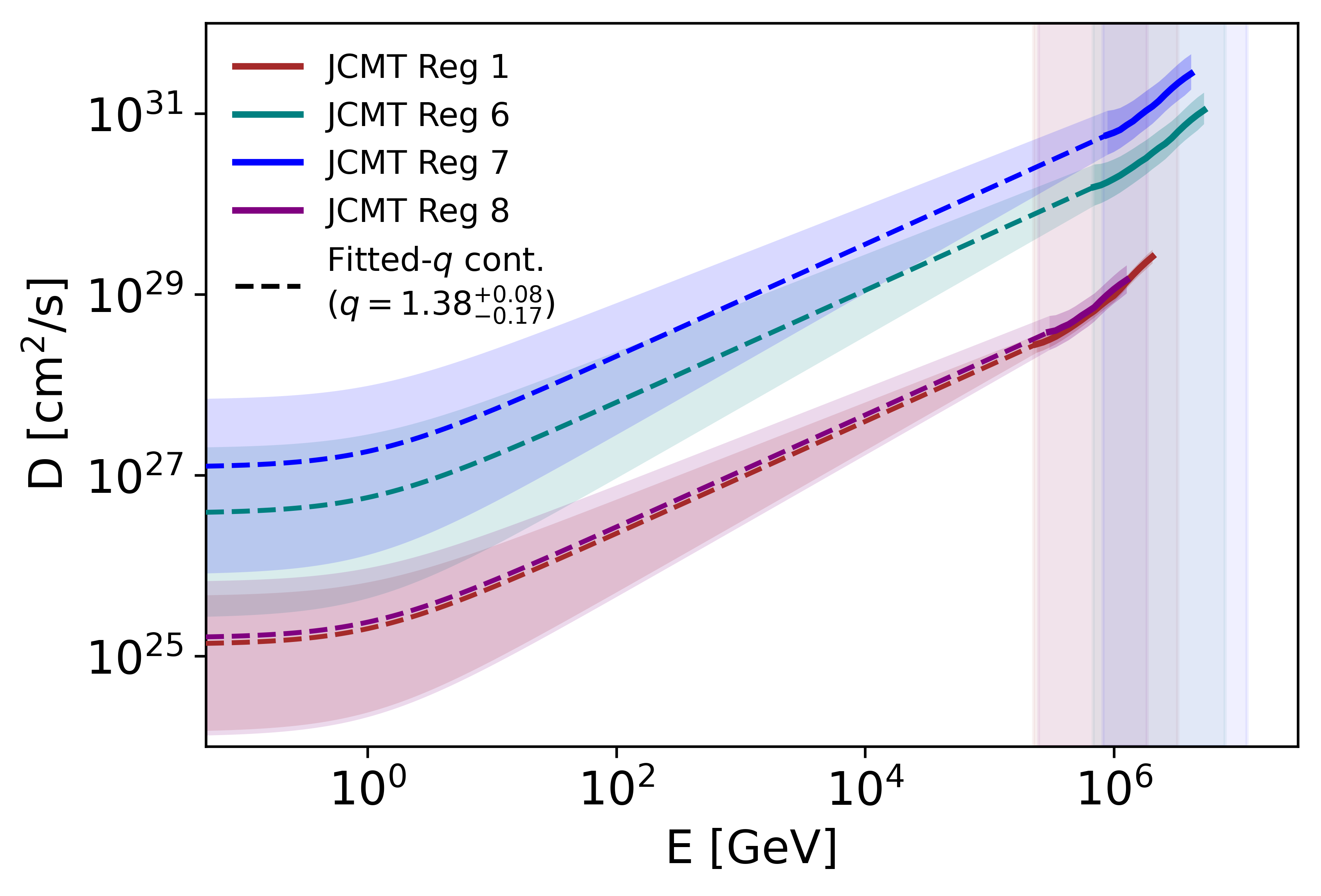}
    \caption{As Fig.~\ref{fig:D_err} (\textit{left}), but with the NGC\,1333
curves continued below the data-informed intervals using the measured
joint index, $q=1.38^{+0.08}_{-0.17}$. Each continuation is anchored
to the measured $D(E)$ at the lower edge of its interval. The shaded
envelopes include the uncertainty in $q$ and the bootstrap
uncertainty at the anchor point. IRAS\,4A is omitted because its
structure functions are saturated at the resolved scales, preventing
a reliable estimate of $q$.} 
    \label{fig:D_err_fitted_q}
\end{figure}

The diffusion coefficients are shown as a function of energy in
Fig.~\ref{fig:D_err}, and representative values within the
data-informed intervals are listed in Table~\ref{tab:diffusion}.
These intervals span 0.24--5.3\,PeV across the NGC\,1333 regions,
and 10--57\,TeV and 12--45\,TeV for IRAS\,4A Regions~1 and~2,
respectively. Their lower edges coincide with the nominal intervals;
their upper edges are lower because the largest separations are
excluded by the lag cap for NGC\,1333 and the trim for IRAS\,4A
(Sec.~\ref{sec:field_structure}). 

Below the data-informed intervals, Fig.~\ref{fig:D_err} shows
continuations using a Kolmogorov index. Fig.~\ref{fig:D_err_fitted_q}
shows the NGC\,1333 continuations using the measured joint index.
We list the resulting $D(E)$ values at 1\,GeV in
Table~\ref{tab:diffusion} for comparison with literature values typically 
quoted at this energy. These values should be regarded as extrapolations rather than direct 
observational constraints. 

\subsubsection{Regional and scale variations}
\label{sec:D_regions}

Within the data-informed intervals, the NGC\,1333 curves have similar
shapes and differ mainly in normalisation. Regions~6 and~7 lie one to
two orders of magnitude above Regions~1 and~8. Region~7, which
contains IRAS~4A, reaches $D \approx 3\times10^{31}\,
{\rm cm^2\,s^{-1}}$ at the upper edge of its interval. Regions~6 and~7
also have the strongest magnetic fields ($B \approx 124$ $\mu$G and
142\,$\mu$G; Table~\ref{tab:para1}). A stronger field places the
data-informed interval at higher energies and, at fixed density and
fluctuation power, gives a larger $D$ (Secs.~\ref{sec:extrapolation_limits}
and~\ref{sec:struc_to_d_e}). All curves steepen towards the upper edge of their intervals
because of the finite range of sampled scales
(Sec.~\ref{sec:extrapolation_limits}); this feature does not
indicate a change in spectral index. 

Across their respective data-informed intervals, the two IRAS~4A
regions give $D \approx (3.5$--$8)\times10^{28}\,
{\rm cm^2\,s^{-1}}$. Their structure functions saturate beyond
$\approx160$\,au (2.5 beams), indicating that much of the fluctuation
power lies on scales close to or below the effective resolution. The
data-informed estimates therefore rely largely on the rise of $S_2$
over the first few bins, and their energy dependence is poorly
constrained, particularly for Region~2. In bootstrap resampling of
Region~2, 85--95 per cent of realisations give a lower $D$ than the
central estimate, while a small fraction give values an order of
magnitude or more higher. The estimate is therefore sensitive to the
vectors included, producing the asymmetric uncertainties listed in
Table~\ref{tab:diffusion}.

The Kolmogorov continuations give $D(1\,\mathrm{GeV}) \approx 5.5\times10^{26}$--$7.2\times10^{28}\,\mathrm{cm^2\,s^{-1}}$ for NGC\,1333 and $\approx 1.1$--$2.7\times10^{27}\,\mathrm{cm^2\,s^{-1}}$ for IRAS\,4A. For NGC\,1333, Fig.~\ref{fig:D_err_fitted_q} also shows continuations using the measured joint index reported in Sec.~\ref{sec:field_structure}. These give $D\propto E^{0.62}$ in the relativistic regime, compared with $D\propto E^{1/3}$ for a Kolmogorov spectrum. The resulting $D(1\,\mathrm{GeV})$ values are lower by $\sim1.5$ dex: approximately $2\times10^{25}\,\mathrm{cm^2\,s^{-1}}$ for Regions 1 and 8, and $\approx6\times10^{26}$--$2\times10^{27}\,\mathrm{cm^2\,s^{-1}}$ for Regions 6 and 7. The uncertainty in the measured index alone spans more than an order of magnitude in $D$ at 1 GeV. These low-energy values therefore depend mainly on the assumed spectrum of unresolved fluctuations, rather than on the measured interval. 

IRAS~4A lies within NGC\,1333 Region~7, but is unresolved in the JCMT
map (Fig.~\ref{fig:reg_pa}). The two datasets therefore probe the same
structure over spatial scales that differ by about two orders of
magnitude. For comparable field strengths ($B \approx 100$--$110$
and 142\,$\mu$G), their data-informed energy intervals lie far apart:
10--57\,TeV for IRAS~4A and 0.88--4.2\,PeV for Region~7 (see Fig.~\ref{fig:D_err}, right panel). A comparison
at the energies sampled by IRAS~4A therefore requires extrapolating the Region~7
curve below its measured interval. At 10--57\,TeV, both Region~7 
continuations give $D \approx 4\times10^{29}$--$2\times10^{30}\,
{\rm cm^2\,s^{-1}}$, one to two orders of magnitude above the IRAS~4A
values. These energies are only one to two decades below the Region~7
interval, and its two continuations differ there by less than a factor
of about four. The comparison is therefore relatively insensitive to
the assumed spectral index. Taken at face value, the lower $D$ and
larger angular dispersion in IRAS~4A
($\sigma_\phi \approx 40$--$48^\circ$, compared with $21^\circ$ in
Region~7) indicate more efficient CR scattering in the IRAS\,4A core than in
the surrounding NGC 1333 cloud region. This suggests that the scattering environment can
vary substantially within a single analysis region, on scales
unresolved by the JCMT map. The IRAS\,4A values are, 
however, only indicative, and confirming this difference will require larger
samples. 

\subsubsection{Dependence on local cloud physical conditions}
\label{sec:dependence_on_cloud_properties}

\begin{table*}[]
\centering
\caption{{Diffusion coefficients for each region. Columns from left to right are the width of the separation bins of the structure function, $\Delta\ell$; the data-informed energy
interval, its logarithmic midpoint, $E^{\rm mid}_{\rm meas}$, and $D$ at that
energy; $D$ at 1\,GeV assuming a Kolmogorov spectral continuation; the spectral index $q$
measured from the direction-balanced structure function; and, for NGC~1333,
$D$ at 1\,GeV from the continuation with the joint index.}}
\label{tab:diffusion}
\renewcommand{\arraystretch}{1.2}
\setlength{\tabcolsep}{5pt}
{
\begin{tabular}{cc|ccccccc}
\hline\hline
 \multicolumn{2}{c|}{Region} & $\Delta\ell$ & $E$-band & $E_{\rm meas}^{\rm mid}$ & $D_{\rm meas,mid}$ & $D_{\rm ext}^{\rm 1 GeV}$ (Kol.) & $q$ & $D_{\rm ext}^{\rm 1 GeV}$ ($q$-fit) \\
 & & (pc) & (TeV) & (TeV) & (cm$^{2}$ s$^{-1}$) & (cm$^{2}$ s$^{-1}$) & & (cm$^{2}$ s$^{-1}$) \\ \hline
\multirow{9}{*}{NGC\,1333} & 1 & 0.012 & 238--2035 & 696 & $(6.60^{+1.28}_{-1.00})\times10^{28}$ & $(5.52^{+1.45}_{-1.06})\times10^{26}$ & 1.20$^{+0.11}_{-0.14}$ & $(2.01^{+4.56}_{-1.78})\times10^{25}$ \\
 & \textit{2} & \multicolumn{7}{c}{\textit{Not measured (3 spectral modes)}} \\
 & \textit{3} & \multicolumn{7}{c}{\textit{Not measured (3 spectral modes)}} \\
 & \textit{4} & \multicolumn{7}{c}{\textit{Not measured (2 spectral modes)}} \\
 & \textit{5} & \multicolumn{7}{c}{\textit{Not measured (2 spectral modes)}} \\
  & 6 & 0.012 & 696--5280 & 1806 & $(3.21^{+1.89}_{-1.07})\times10^{30}$ & $(2.09^{+1.62}_{-0.78})\times10^{28}$ & 1.60$^{+0.32}_{-0.32}$ & $(5.68^{+22.77}_{-5.24})\times10^{26}$ \\ 
 & 7 & 0.012 & 884--4160 & 1806 & $(1.07^{+0.71}_{-0.39})\times10^{31}$ & $(7.20^{+6.18}_{-2.84})\times10^{28}$ & 1.97$^{+0.45}_{-0.56}$ & $(1.84^{+7.96}_{-1.70})\times10^{27}$ \\ 
 & 8 & 0.012 & 302--1263 & 618 & $(6.48^{+2.77}_{-2.04})\times10^{28}$ & $(6.95^{+3.52}_{-2.55})\times10^{26}$ & 1.87$^{+0.56}_{-0.59}$ & $(2.38^{+7.06}_{-2.16})\times10^{25}$ \\ 
 & \textit{9} & \multicolumn{7}{c}{\textit{Not measured (2 spectral modes)}} \\
\hline
\multirow{2}{*}{IRAS\,4A} & 1 & 0.0002 & 10--57 & 22 & $(3.47^{+1.00}_{-0.78})\times10^{28}$ & $(1.09^{+0.47}_{-0.29})\times10^{27}$ & \textit{1.08$^{+0.21}_{-0.20}$}$^{\star}$ & -- \\
 & 2 & 0.0002 & 12--45 & 22 & $(7.74^{+2.93}_{-4.35})\times10^{28}$ & $(2.70^{+1.42}_{-1.63})\times10^{27}$ & \textit{1.43$^{+1.11}_{-1.10}$}$^{\star}$ & -- \\ 
\hline
\end{tabular}}
\tablefoot{{Uncertainties are the bootstrap bounds described in Appendix~\ref{app:uncertainty}, each enclosing 68 per cent of the realisations
on its side of the central value. $\Delta\ell$ is the narrowest bin width, to a precision of 0.2 pixel, for which every separation bin contains at least three pairs. It is determined after the lag cap (NGC\,1333) or after the trim and re-binning (IRAS\,4A), and sets the low-energy edge of the data-informed interval through $\ell_{\rm min} = 2\Delta\ell$ (Eq.~\ref{eq:e_band_low}). $E^{\rm mid}_{\rm meas}$ is the logarithmic
midpoint of the data-informed interval, on the energy grid of the calculation
(0.05 dex spacing). The data-informed intervals are narrower than the nominal intervals indicated by Eqs.~\ref{eq:e_band_low} and~\ref{eq:e_band_high} because
separations beyond the lag cap (NGC\,1333) or in the large-separation structure downturn
(IRAS\,4A) are excluded (Appendix~\ref{sec:modes}). NGC\,1333 Regions 2-5 and 9 retain three or fewer spectral modes after the lag cap and are not
analysed further. The structure-function slopes, $q-1$, differ from zero at
only 1.5--1.9$\sigma$ for the individual NGC\,1333 regions. Their indices are
consistent with the joint index, $q = 1.38^{+0.08}_{-0.17}$ (3.4$\sigma$),
which is used for the fitted-$q$ continuation (final column). \\
$^{\star}$ For IRAS\,4A, the structure functions are saturated at the resolved scales (i.e. their slopes are consistent with zero), so no spectral index is measured. The values are shown above for reference only. For IRAS\,4A Region 2, 85-95 per cent of the bootstrap realisations fall below the measured value of $D$, so its bounds are strongly asymmetric.
}}
\end{table*}

\begin{figure}
    \centering
    \includegraphics[width=\linewidth]{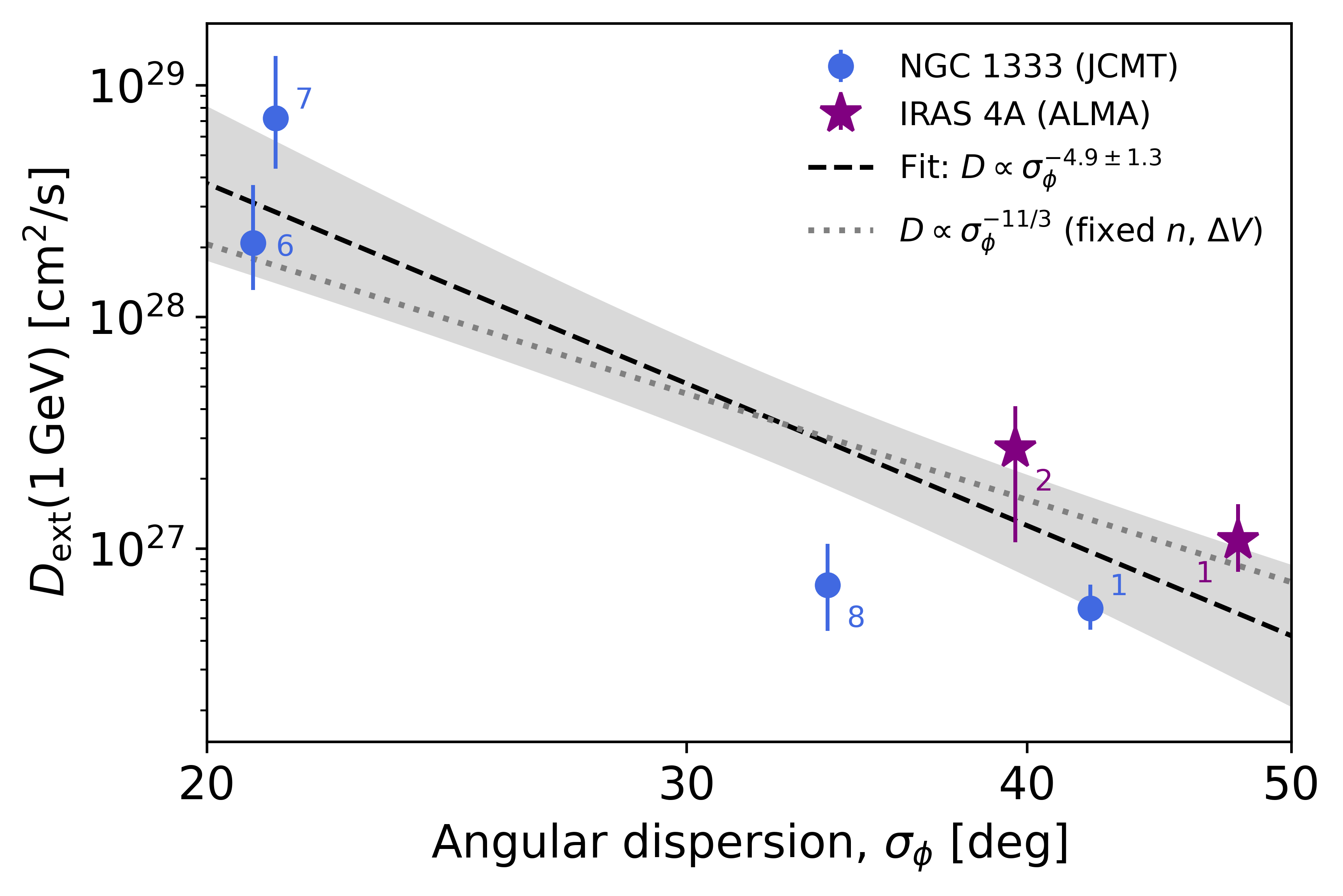} 
\caption{{Diffusion coefficient at 1 GeV from the Kolmogorov continuation
against the angular dispersion, $\sigma_\phi$, for the NGC~1333 (JCMT; 
circles) and IRAS~4A (ALMA; stars) regions, labelled by region number, with 
bootstrap uncertainties (Appendix~\ref{app:uncertainty}). The dashed line and grey band show
the unweighted least-squares power-law fit. The dotted line shows the
expected $D\propto\sigma_\phi^{-11/3}$ scaling for regions with the
same density, linewidth, spectral shape, and resolution, normalised at
the geometric mean of the points.}}
    \label{fig:ang_dis}
\end{figure}

The Kolmogorov-extended values of $D(1\,{\rm GeV})$ show an
anti-correlation with the magnetic-field angular dispersion
(Fig.~\ref{fig:ang_dis}). This is expected from the method. At a 
fixed resonant scale, $D \propto B^2\rho^{1/2}/\mathcal{P}$, where
$\mathcal{P}$ is the fluctuation-power amplitude. The DCF estimate
gives $B \propto n^{1/2}\Delta V\sigma_\phi^{-1}$, while for a fixed
spectral shape $\mathcal{P}\propto\sigma_\phi^2$. For the
Kolmogorov continuations, the lower-energy anchor is proportional to
$B$ and $D\propto E^{1/3}$. Combining these scalings gives $D(1\,{\rm GeV}) \propto
n^{4/3}\Delta V^{5/3}\sigma_\phi^{-11/3}$ for regions with the same spectral shape and resolution. 

The four NGC\,1333 regions have similar measured spectral
shapes and the same separation bins. Their extrapolated
$D(1\,{\rm GeV})$ values broadly follow the combined dependence
on density, linewidth, and angular dispersion, and a fit of the
inferred values against those predicted by this scaling has
a slope of $1.16\pm0.13$. Fig.~\ref{fig:ang_dis} instead shows the dependence on angular
dispersion alone. Across both datasets, the fitted relation is
$D\propto\sigma_\phi^{-4.9\pm1.3}$. NGC\,1333 Region~7 lies
approximately 2.5 times above this fit, while Region~8 lies
approximately four times below it. These offsets show that
angular dispersion alone does not account for all the regional
variation. 
Within
NGC\,1333, regions with smaller $\sigma_\phi$ also have higher
densities and linewidths, steepening the relation relative to
$\sigma_\phi^{-11/3}$. The IRAS\,4A regions lie above Regions~1 and~8
in Fig.~\ref{fig:ang_dis}; they have similar angular dispersions but
higher densities and linewidths. The observed relation is therefore
a consistency check of the method, rather than an independent
empirical law.

The angular dispersion includes both small-scale fluctuations and
large-scale field morphology. In the NGC\,1333 regions, smooth
curvature of the ordered field contributes substantially to
$\sigma_\phi$ (Appendix~\ref{app:limitations}). The correlation with
$D$ therefore reflects field irregularities across the range of
scales relevant to CR transport. Comparisons of $D$ with density,
linewidth, and field strength show related trends within NGC\,1333;
the IRAS\,4A regions lie below those NGC\,1333 trends
(Appendix~\ref{app:D_vs_para}). 

\section{Discussion}
\label{sec:discussions}

\subsection{Interpretation and applicability of the diffusion coefficient}
\label{sec:d_e_shape}

\subsubsection{Energy dependence and data-informed range}
\label{sec:d_e_physical_features} 

\begin{figure*}
    \centering
    \includegraphics[width=0.7\textwidth]{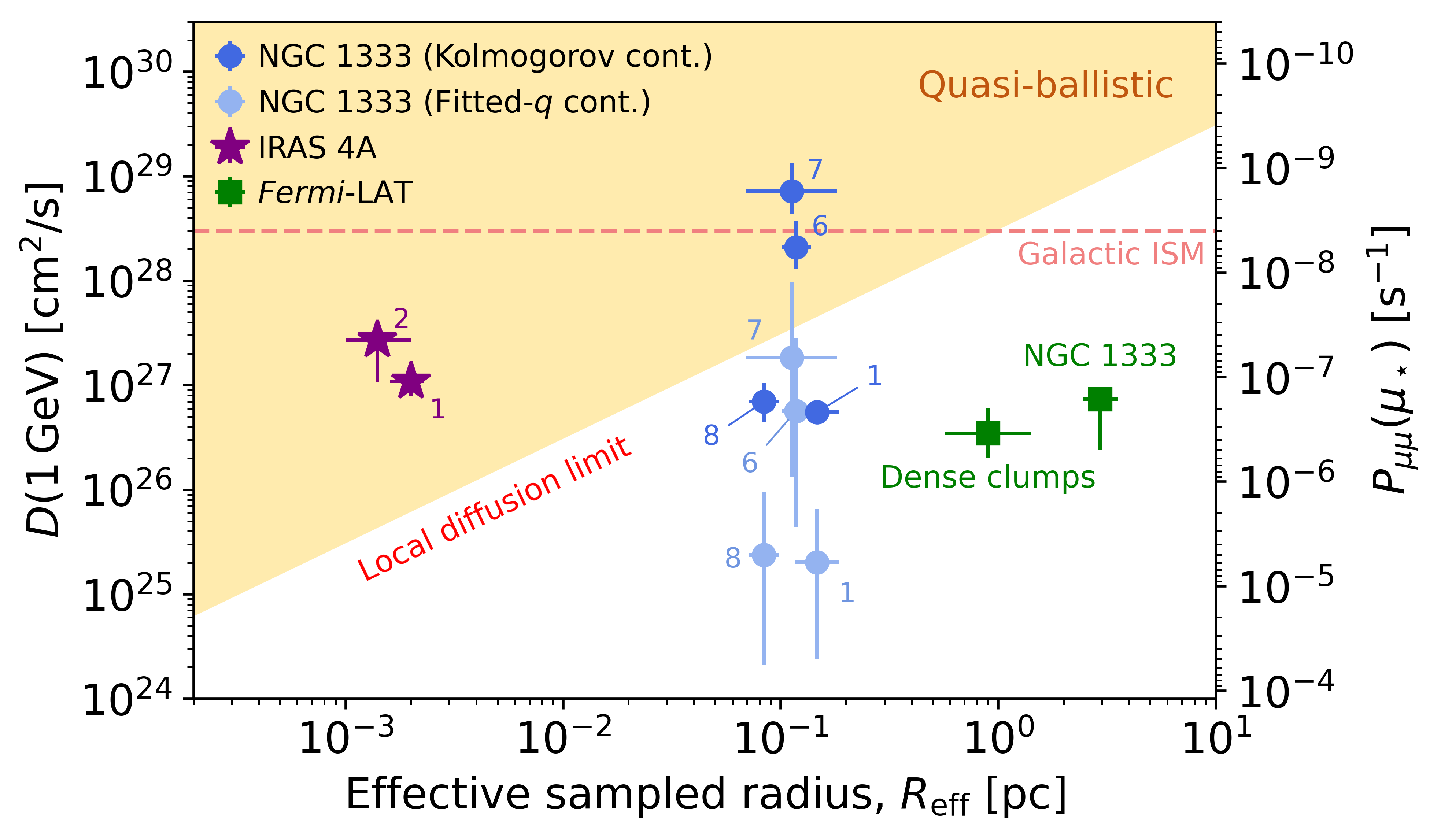} 
    \caption{Diffusion coefficient at 1\,GeV against the characteristic
spatial scale of each measurement. For our regions, this scale
is the effective radius, $R_{\rm eff}$
(Table~\ref{tab:para1}); horizontal bars span the fitted
semi-minor and semi-major axes, $b$ and $a$.
Dark-blue circles and purple stars show the Kolmogorov-extrapolated
values for NGC\,1333 and IRAS\,4A, respectively. Light-blue circles
show the NGC\,1333 continuations using the measured joint index.
Points are labelled by region number. Green squares show the
\textit{Fermi}-LAT estimates of \citet{Jiang2025MNRAS} and
\citet{Yang2023NatAs}; the spatial scales and vertical bars are
described in the text. The right-hand axis converts $D$ to the
equivalent $P_{\mu\mu}(\mu_\star)$ using Eq.~\ref{eq:D}.
The light-red dashed line marks a typical Galactic value at
GeV energies \citep[e.g.][]{Strong2007ARNPS}.
Above the limiting coefficient in Eq.~\ref{eq:Dlim}
(shaded region), the scattering mean free path exceeds the
plotted spatial scale.} 
    \label{fig:D_vs_size}
\end{figure*}

The shape of the inferred $D(E)$ curve can be understood from the dominant energy dependencies in
Eqs.~\ref{eq:k_0} and~\ref{eq:j_def}-\ref{eq:D}. 
Combining Eqs.~\ref{eq:pmumu} and~\ref{eq:D}, and retaining only
the terms that depend on CR energy, gives
$D(E) \propto {K_0}/{(\mathcal{J}(K_0)\,\omega_{\rm L}^2(E))}$.
At fixed pitch angle, $K_0\propto\omega_{\rm L}$. For CR protons,
$\omega_{\rm L}\propto1/\gamma$, where
$\gamma=E/(m_{\rm p}c^2)+1$. In the relativistic limit,
$\gamma\propto E$, so $K_0\propto E^{-1}$ and
$D(E)\propto E/\mathcal{J}(K_0)$. Bends and other features in
$D(E)$ therefore trace how the resonant wavelength,
$2\pi/K_0$, moves through the measured magnetic fluctuation spectrum.

If the fluctuation spectrum follows a power law,
$\mathcal{P}_\parallel(k)\propto k^{-q}$, across the wavenumber range
contributing to $\mathcal{J}(K_0)$, then $D(E)\propto E^{2-q}$ in the
relativistic regime, provided the resonant scale lies well within that
range. Our spectra contain a limited number of modes and cover a finite
range of scales. Near the upper edge of the data-informed intervals,
this finite range affects $\mathcal{J}(K_0)$ and steepens $D(E)$
(Sec.~\ref{sec:D_regions}). The slopes of the resulting curves
therefore cannot be used to infer $q$ directly. We instead estimate
$q$ from the scale dependence of $S_2(\ell)$
(Appendix~\ref{sec:modes}).

The index $q$ measured for the NGC~1333 regions can be compared with the velocity structure of
molecular clouds. If magnetic-field fluctuations follow velocity
fluctuations, as in Alfv\'enic turbulence, a linewidth--size relation
$\sigma_v\propto L^\beta$ implies
$S_2\propto\ell^{2\beta}$ and hence $q=1+2\beta$~\citep{Larson1981MNRAS}. The cloud-scale
value $\beta\approx0.5$--$0.7$~\citep{Solomon1987ApJ,
Heyer2004ApJ} gives $q\approx2$--$2.4$, substantially steeper than the
spectrum inferred for NGC\,1333. The joint 
NGC\,1333 index (Sec.~\ref{sec:field_structure}) instead corresponds to $\beta\approx0.2$ and implies
more field structure on small scales than a Larson-type cascade
would predict. 

At core scales, IRAS\,4A provides a complementary comparison. It has a larger angular dispersion than its
host Region~7 (Table~\ref{tab:para1}) and a lower inferred $D$ when 
the two are compared at common energies using the continuation
described above (Sec.~\ref{sec:D_regions}). Its larger angular dispersion includes the curvature of the
hourglass field, so this comparison does not directly measure
how turbulent fluctuation amplitudes vary with scale.
Nevertheless, the difference in magnetic structure is
consistent with the breakdown of Larson relations in dense cores, where
turbulence becomes subsonic and coherent
\citep[e.g.][]{Goodman1998ApJ, Pineda2010ApJ}\footnote{We note that these studies concern starless cores, at an
earlier evolutionary stage than IRAS\,4A.}, and with field
pinching during gravitational collapse, as indicated by the
hourglass morphology of IRAS\,4A~\citep{Girart2006Sci}. These results caution against using Larson-type scalings
alone to extrapolate CR scattering rates from cloud to
core scales. 

Below the directly probed range, $D(E)$ depends on the assumed spectrum of unresolved fluctuations. We use a Kolmogorov continuation as a reference for comparison with other works. Using the measured joint index for NGC\,1333 instead lowers the inferred values at 1\,GeV by $\sim1.5$ dex (Sec.~\ref{sec:D_regions}). The 1\,GeV coefficients are therefore extrapolations whose values depend strongly on the assumed spectrum. 

The extrapolated $D(E)$ curves also flatten at low energies
within the adopted prescription. As $\gamma\to1$, the
gyro-frequency and $K_0$ approach constant values in the
expressions used here. This behaviour persists when the
fluctuation spectrum is extrapolated, distinguishing it from
the plateau caused by truncating the measured spectrum at 
$\ell_{\rm min}$. However, interpreting it in the
sub-relativistic regime requires retaining the actual
particle speed in both the resonance mapping and the
diffusion calculation. 

\subsubsection{Scale of applicability of the diffusion coefficient}
\label{sec:validity_energy_range}

The inferred $D(E)$ describes diffusive transport only on scales
that are much larger than the corresponding parallel scattering
mean free path, 
\begin{equation}
    \ell_{\rm diff} = \frac{3D}{v} \simeq \frac{3D}{c}
    \label{eq:mfp}
\end{equation}
for relativistic particles. The diffusion approximation requires
$L\gg\ell_{\rm diff}$, so that particles scatter many times over
the relevant transport scale $L$. On scales
$L\lesssim\ell_{\rm diff}$, this condition fails and transport
can be quasi-ballistic. The condition
$\ell_{\rm diff}=L$ defines a characteristic limiting coefficient,
\begin{equation}
D_{\rm lim} \equiv \frac{Lc}{3}
\simeq 3\times10^{28}
\left(\frac{L}{1\,{\rm pc}}\right)
\,{\rm cm^2\,s^{-1}} \ .
\label{eq:Dlim}
\end{equation}
A diffusive description on scale $L$ requires
$D(E)\ll D_{\rm lim}$. This limit depends only on $L$ and the microscopic speed of the CR
particles, $v\simeq c$. It is independent of the normalisation of
our inferred CR scattering rate.  

To compare this limit with our results, we take $L=R_{\rm eff}$. The analysed regions have
$R_{\rm eff}\sim10^{-3}\,{\rm pc}$ for IRAS\,4A and
$\sim0.05$--$0.15\,{\rm pc}$ for NGC\,1333.
These sizes give $D_{\rm lim}\sim10^{25}$--$10^{27}
\,{\rm cm^2\,s^{-1}}$.
Our inferred coefficients exceed these limits throughout the
directly probed TeV--PeV intervals by a margin that increases with energy. A diffusive description of CR transport within an
individual analysis region therefore does not apply at these energies. The observations instead constrain the local pitch-angle scattering
coefficient, $P_{\mu\mu}$, through the magnetic-fluctuation power
resonant with each CR energy. This is a microphysical quantity. Equation~\ref{eq:D} expresses this
scattering coefficient as the $D(E)$ that would describe transport
if the CR particles sampled the same scattering conditions over scales
$L\gg\ell_{\rm diff}$. The reported values therefore characterise
the local scattering environment and its variation between regions.
Transport over larger scales depends on how CRs sample regions
with different scattering rates. 

\subsection{Comparison with other CR transport constraints}
\label{sec:comparison_prev_results}

Our data-informed results apply to CR protons at approximately
10\,TeV to 5\,PeV, whose resonant magnetic fluctuations are sampled
by the ALMA and JCMT observations. Most existing CR propagation
constraints apply at MeV--TeV energies and are inferred from CR
densities or emission morphologies averaged over pc to Galactic
scales. Direct comparisons therefore require overlapping CR energies
and must account for the different spatial scales over which CR 
transport is constrained.  

Hadronic $\gamma$-ray observations offer a way to extend these
comparisons to the energies sampled here. CR protons produce
$\gamma$-rays with characteristic energies of about a tenth of
their own through neutral-pion decay when they undergo a hadronic interaction~\citep{Kelner2006PhRvD}. 
Our data-informed intervals therefore correspond approximately 
to $\gamma$-ray energies of 1--6\,TeV for IRAS\,4A and
25--500\,TeV for NGC\,1333. These energies are accessible to
current and forthcoming observatories. The Cherenkov Telescope
Array Observatory (CTAO) is designed to cover approximately
20\,GeV--300\,TeV \citep{CTAO2019book}, while air-shower arrays such as the Large High Altitude Air Shower Observatory, LHAASO \citep{Cao2021Natur} reach the PeV range. Where hadronic emission
from a cloud is detectable, these observations can constrain
CR transport at energies overlapping our analysis.

The two approaches probe transport on different spatial scales.
The sub-mm observations constrain local scattering in individual
structures, with substantial variation even within a single
clump (Sec.~\ref{sec:D_regions}). The $\gamma$-ray observations
constrain effective transport over larger regions containing
many such structures. Observations of the same system at
overlapping CR energies could therefore test whether, and on
what scales, diffusion emerges from locally varying scattering.
They could also establish how the resulting diffusion coefficient
depends on the local rates, including whether the most strongly
scattering structures control the transport
\citep[e.g.][]{Ewart2026MNRAS}. 

For comparison with existing constraints, we use the extrapolated
values at 1\,GeV listed in Table~\ref{tab:diffusion}.
Fig.~\ref{fig:D_vs_size} shows both the Kolmogorov continuations
and, for NGC\,1333, those using the measured joint index.
The figure compares these values with literature estimates as
a function of the characteristic spatial scale sampled by each
measurement. 

For \citet{Jiang2025MNRAS}, the plotted scale is the size of
their analysis region. For \citet{Yang2023NatAs}, we use the
size of the NGC\,1333 clump in their sample. Their diffusion
constraint is inferred from the stacked spectrum of five dense
clumps in Taurus and Perseus, and the plotted range covers
assumed energy indices of 0--1. The spatial scale therefore
identifies the region over which the magnetic structure or CR
distribution is constrained. It does not establish that the
inferred coefficient describes diffusion within that region
(Sec.~\ref{sec:validity_energy_range}). 

The
vertical bars are also not strictly comparable. For our regions, they show bootstrap uncertainties, including
the 68 per cent range of $q$ for the continuations using the
joint index. For \citet{Jiang2025MNRAS}, they show the quoted
uncertainty, whereas for \citet{Yang2023NatAs}, they span the
assumed energy indices. The figure therefore compares the
inferred coefficients across observational scales, subject
to differences in the methods, spectral assumptions, and
uncertainty estimates. 

\subsubsection{NGC\,1333 and its host cloud environment}
\label{sec:host_cloud}

NGC\,1333 allows independent CR diagnostics to be compared within
the same molecular-cloud complex. \textit{Fermi}-LAT observations
of dense clumps in Taurus and Perseus, including NGC\,1333, showed 
a deficit of low-energy $\gamma$-ray emission relative to that
expected for uniform CR penetration \citep{Yang2023NatAs}.
Modelling this deficit as energy-dependent CR shielding gave 
$D(1\,{\rm GeV})\simeq(2$--$6)\times10^{26}
\,{\rm cm^2\,s^{-1}}$ on approximately pc scales.
A subsequent analysis of Perseus found spatial variations in the
GeV CR spectrum over several parsecs, including a $\gamma$-ray
deficit toward NGC\,1333 below approximately 2\,GeV
\citep{Jiang2025MNRAS}. Their inferred coefficient was 
$D(1\,{\rm GeV})\approx7\times10^{26}\,{\rm cm^2\,s^{-1}}$.
For comparison, our values at 1\,GeV span approximately
$5.5\times10^{26}$--$7.2\times10^{28}$ cm$^2$ s$^{-1}$ with the Kolmogorov
continuation and $2\times10^{25}$--$1.8\times10^{27}$ cm$^2$ s$^{-1}$ for
NGC\,1333 with the measured index (Sec.~\ref{sec:D_regions}), so the 
choice of continuation limits how quantitative a comparison at this energy 
can be. 

Nevertheless, the local values of $D(E)$ vary by about two orders of magnitude 
between sub-pc regions of NGC\,1333. The \textit{Fermi}-LAT
estimates lie near the lower end of the Kolmogorov range,
close to the values for Regions~1 and~8
(Fig.~\ref{fig:D_vs_size}). This is consistent with transport
through a heterogeneous medium in which strongly scattering
structures exert disproportionate control over the effective
coefficient on larger scales \citep{Ewart2026MNRAS}.
The comparison between IRAS\,4A and its surrounding Region~7
also suggests that more strongly scattering structures occur
within the cloud (Sec.~\ref{sec:D_regions}). However, the
IRAS\,4A values are indicative, and the comparison requires
extrapolating the Region~7 spectrum. These results therefore
suggest a possible connection between local scattering and
pc-scale transport, but do not establish which structures
control the cloud-scale coefficient. 

Lower-energy CRs are constrained independently by ionization
measurements. \citet{Luo2026arXiv} infer
$\zeta\simeq(0.4$--$2)\times10^{-17}\,{\rm s^{-1}}$
from H\,I narrow self-absorption, while
\citet{Pineda2024A&A} obtain typical values of
$\zeta\approx10^{-16.5}\,{\rm s^{-1}}$ from
DCO$^+$/H$^{13}$CO$^+$. At representative NGC\,1333 columns,
the ionization response is dominated by protons near 10\,MeV.
These observations therefore probe a different part of the CR
population from both the \textit{Fermi}-LAT measurements and
our TeV--PeV analysis. Spatial variations inferred from the
deuterated tracers also carry additional uncertainties in
heated and shocked gas. We therefore do not use these
constraints quantitatively here
(Appendix~\ref{app:low_energy_measurements}).  

\subsubsection{CR transport in other environments}
\label{sec:broader_comparison}

Other molecular clouds provide complementary examples of spatially
and energy-dependent CR transport. Passive clouds are broadly
consistent with irradiation by the Galactic CR sea
\citep{Aharonian2020PhRvD,Peron2022A&A}.
Clouds near CR accelerators receive enhanced irradiation, but
the CR population reaching their interiors also depends on the
local transport conditions. In RX~J1713.7$-$3946, for example,
shock--clump interactions can amplify magnetic turbulence around
dense clumps. The resulting increase in scattering inhibits CR
penetration, particularly at lower energies
\citep{Inoue2012ApJ,Sano2013ApJ,Sano2020ApJ,
Gabici2014MNRAS,Celli2019MNRAS}.
CR-generated fluctuations and pre-existing layers of slow
diffusion can similarly suppress low-energy CR penetration
\citep{Inoue2019ApJ,Ng2026PASJ}.
These observations constrain the CR population that reaches
the gas. Our method provides a complementary constraint on the
magnetic fluctuations regulating that access. 

On larger scales, approximately $1/r$ CR-density profiles around
young stellar clusters and in the Galactic Centre have been
interpreted in terms of approximately homogeneous diffusion over
scales from pc to hundreds of pc
\citep{Aharonian2019NatAs,HESS2016Natur,Devin2025arXiv}.
Such profiles need not imply uniform scattering on smaller
scales. Regional variations like those inferred by our results could coexist
with an approximately homogeneous diffusion coefficient on 
larger scales, depending on how CRs sample the local scattering
environments \citep{Ewart2026MNRAS}. 
Further examples, including comparisons with TeV halos and
other star-forming regions, are discussed in
Appendix~\ref{app:comparison}.  

\section{Summary and conclusions}
\label{sec:conclusions}

In this work, we used a Fourier-space scattering framework to infer local CR
pitch-angle scattering rates and effective diffusion coefficients
from JCMT polarisation observations of NGC\,1333 and ALMA
observations of IRAS\,4A. The calculation combines the estimated
magnetic-field strengths with the measured fluctuation spectra.
The framework applies to charged CR particles, with the resonant
scale determined by their rigidity. The retained fluctuations
probe rigidity intervals corresponding to proton energies of
approximately 10--57\,TeV in IRAS\,4A and 0.24--5.3\,PeV across
NGC\,1333. Our main findings are: 
\begin{enumerate}
\item The inferred $D(E)$ varies substantially between regions.
    At the logarithmic midpoints of the data-informed intervals,
    the coefficients span approximately
    $3\times10^{28}$--$10^{31}\,{\rm cm^2\,s^{-1}}$
    (Sec.~\ref{sec:D_regions}; Table~\ref{tab:diffusion}).
    The Kolmogorov-extrapolated values at 1\,GeV decrease with
    increasing magnetic-field angular dispersion.
    This relation follows from the dependence of the method
    on angular dispersion, density, and linewidth, and provides
    a consistency check rather than an independent empirical law
    (Sec.~\ref{sec:dependence_on_cloud_properties}).
    \item  The four analysed NGC\,1333 regions are consistent with a
    shared fluctuation-spectrum index,
    $q=1.38^{+0.08}_{-0.17}$, flatter than the Kolmogorov value.
    The IRAS\,4A structure functions are saturated at the
    resolved scales, preventing a reliable index measurement
    (Sec.~\ref{sec:field_structure}).
    The Kolmogorov continuations give reference values at
    1\,GeV of approximately
    $6\times10^{26}$--$7\times10^{28}\,{\rm cm^2\,s^{-1}}$
    across the two datasets. Using the measured joint index
    for NGC\,1333 lowers its extrapolated values by
    $\sim1.5$\,dex. Comparisons at GeV energies therefore
    depend strongly on the assumed spectrum of unresolved
    fluctuations (Sec.~\ref{sec:D_regions}).
    \item Comparing IRAS\,4A with its surrounding NGC\,1333 Region~7
    at common CR energies suggests scattering that is stronger
    by one to two orders of magnitude in the embedded core
    (Sec.~\ref{sec:D_regions}). This comparison requires
    extrapolating the Region~7 spectrum, and the IRAS\,4A
    coefficients are only indicative.
    On pc scales, the coefficient inferred from
    \textit{Fermi}-LAT observations lies near the lower end
    of our Kolmogorov-extrapolated local values
    (Sec.~\ref{sec:host_cloud}; Fig.~\ref{fig:D_vs_size}).
    Together, these comparisons suggest substantial variation
    in the scattering environment within the cloud. 
    \item Within the data-informed intervals, the inferred scattering
mean free paths exceed the sizes of the individual analysis
regions. The reported coefficients therefore characterise
local scattering conditions and cannot describe diffusion
within those regions
(Sec.~\ref{sec:validity_energy_range}).
A macroscopic diffusive regime, if established, requires
CRs to sample multiple regions with different scattering
rates. Its effective coefficient would depend on how
particles traverse these structures, potentially giving
disproportionate weight to the most strongly scattering
regions (Sec.~\ref{sec:comparison_prev_results}).
Hadronic $\gamma$-ray emission at energies accessible to
current and forthcoming observatories probes CR energies
overlapping our data-informed intervals. Combined
$\gamma$-ray and sub-mm observations could therefore compare
local scattering with transport over larger scales, testing
whether, and on what scales, an effective diffusion
coefficient emerges. 
\end{enumerate}

While the analysis framework presented in this work 
is subject to methodological and data limitations (see Appendix~\ref{app:limitations}), it 
provides a practical route for connecting effective CR transport prescriptions employed on cloud and galaxy scales~\citep[e.g.][]{Strong2007ARNPS, Yang2023NatAs, Ng2026PASJ}  to the scattering microphysics that those prescriptions seek to represent. 
Our results reveal substantial spatial variation in the
inferred local scattering rate. Establishing how these
variations combine into an effective cloud-scale diffusion
coefficient requires modelling CR transport through the
measured magnetic structure. 

\begin{acknowledgements}

STC acknowledges funding by the Helga-Kersten program supporting faculty-specific gender equality targets at Friedrich-Alexander University Erlangen-N\"{u}rnberg (FAU), which supported an internship at the Erlangen Centre for Astroparticle Physics (ECAP) during 2025 in the group of AM. ERO is supported by the RIKEN Special Postdoctoral Researcher Program for junior researchers, the RIKEN Incentive Research Project (``Diffusion and beyond: probing anomalous cosmic-ray transport in magnetized 
molecular clouds''), and also acknowledges support from a Postdoctoral Fellowship of the Japan Society for the Promotion of Science (JSPS KAKENHI Grant No. JP22F22327), hosted by the University of Osaka, during which part of this work was completed. The early stages of this project were initiated when ERO was at the Center for Informatics and Computation in Astronomy (CICA). CICA is hosted by National Tsing Hua University and supported by a grant from the Ministry of Education of Taiwan. 
AM acknowledges support by the German {Deut\-sche For\-schungs\-ge\-mein\-schaft, DFG\/}, project number 452934793. SPL and STC acknowledge support from the National Science and Technology Council (NSTC) of Taiwan under grants 112-2112-M-007-011, 113-2112-M-007-004, and 114-2112-M-007-001. SPL also acknowledges support from the NSTC under grants 115-2112-M-007-017-MY3 and 115-2918-I-007-013. 
The authors are grateful for insightful comments from Tim Linden (Stockholm University) and Dimitry Malyshev (ECAP), which led to refinements in our analysis methodology, and discussion of the limitations of the approach presented in this work, Nami Sakai and the RIKEN Star and Planetary Formation Laboratory, Hidetoshi Sano (Gifu University) and Shigeo S. Kimura (Tohoku University) for comments that broadened our discussion of the implications of this work, and led to improvements in our data analysis approach. This study further benefited from discussions and comments received at the workshop \textit{Cosmic Rays 4: The Salt of the Star-Formation Recipe} (Pisa, Italy, 2026). ChatGPT-5 (OpenAI 2025) and Claude-Opus-4.7 (Anthropic 2026) were used to improve the formatting of Figures 3, 4, 5, A.1., A.2., and B.1., to refine the wording of some parts of the text, and improve the typesetting of Tables 1 and 2. All content where these tools were used was carefully reviewed, edited, and verified for accuracy by the authors, who assume full responsibility for the outcome. 

\end{acknowledgements}

\bibliographystyle{aa}
\bibliography{ref}

\appendix

\section{Supplementary results, methodology and diagnostics}
\label{app:results}

\subsection{Magnetic-field strength estimation}
\label{app:field_strength}

We derive the quantities entering the DCF magnetic-field
estimate (Eq.~\ref{equ:DCF}) from the dust continuum,
polarisation, and NH$_3$ observations. We obtain H$_2$
column-density maps from the continuum intensity, assuming
optically thin dust emission and using the Planck function
at the adopted dust temperature. We assume a line-of-sight
depth of $L=0.1$\,pc, consistent with the characteristic
width of nearby filaments ($d<500$~pc; \citealp{Arzoumanian2011}), and
estimate the volume density as $n({\rm H_2}) = {N({\rm H_2})}/{L}$. 

We calculate the angular dispersion, $\sigma_\phi$, from the
field orientations within each region. We use doubled angles to account for
the $180^\circ$ ambiguity of the polarisation orientations, and subtract
the contribution from measurement uncertainties in quadrature, $\sigma_\phi^2 =
    \sigma_{\rm obs}^2-\left\langle\sigma_{\rm err}^2\right\rangle$, where $\sigma_{\phi,{\rm obs}}$ is the observed angular
dispersion and $\sigma_{\phi,{\rm err}}$ is the uncertainty
in each orientation. 
The same definition is used for the JCMT and ALMA datasets. The resulting dispersion includes
any coherent curvature of the ordered field within the
region. This contribution can cause the DCF field strength
to be underestimated by up to approximately a factor of two
(Appendix~\ref{app:limitations}). 

The velocity dispersion is obtained from the NH$_3$(1,1) data by fitting
the hyperfine structure with the Python package {\tt Pyspeckit}
\citep{Ginsburg2011,Ginsburg2022}. The fitted dispersion, $\sigma_{v,{\rm obs}}$, gives the
FWHM linewidth, 
\begin{equation}
    \Delta V = \sqrt{8\ln 2}\,\sigma_{\rm obs}.
\end{equation}
We subtract the thermal contribution in quadrature to obtain
the non-thermal linewidth, $\Delta V_{\rm nt}$, with the thermal component calculated
from the dust temperature and the molecular mass of NH$_3$. This assumes that the gas and dust temperatures are equal. 

We use the regional mean density and non-thermal linewidth,
together with the regional angular dispersion, in
Eq.~\ref{equ:DCF}. These quantities and the resulting
plane-of-sky magnetic-field strengths are listed in
Table~\ref{tab:para1}.

\subsection{Magnetic-field fluctuation power spectra and mode selection}
\label{sec:modes}

\begin{figure*}
    \centering
    \includegraphics[width=\linewidth]{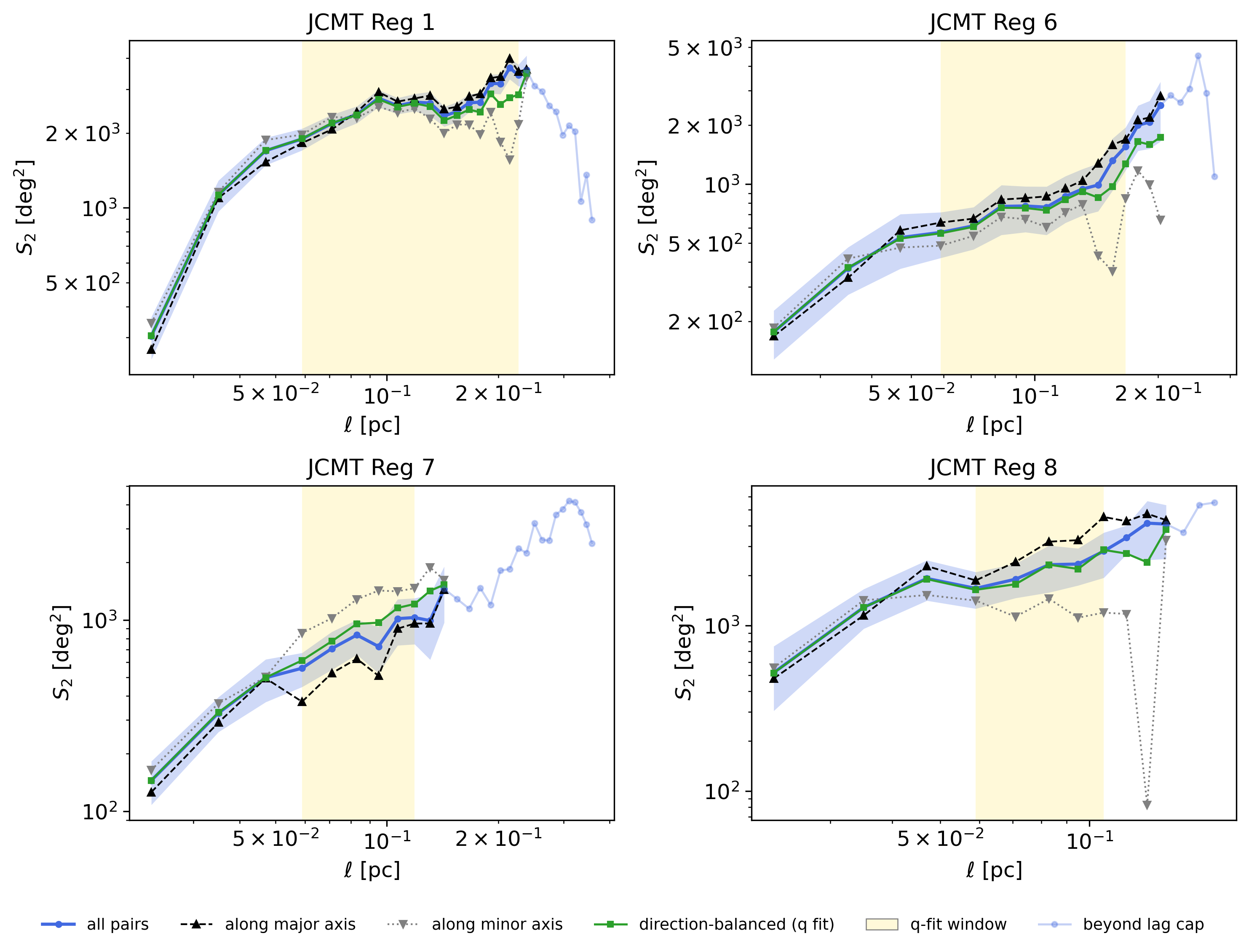}
    \caption{Second-order structure functions, $S_2(\ell)$, of
the four analysed NGC\,1333 regions. Blue circles and lines
show the estimates using all pair directions; shading spans
the 16th--84th percentiles of the bootstrap realisations.
Black dashed lines with upward triangles and grey dotted
lines with downward triangles show the estimates for pairs
oriented along the major and minor axes, respectively.
Green squares show the direction-balanced estimates used
to infer $q$. Yellow shading marks the fitting interval:
separations of at least 2.5 beam FWHM with at least 10
pairs in each direction. Faded points show separations
beyond the lag cap, $2b$, which are excluded from the
analysis.} 
    \label{fig:S_2_functions_NGC1333}
\end{figure*}

\begin{figure*}
    \centering
    \includegraphics[width=\linewidth]{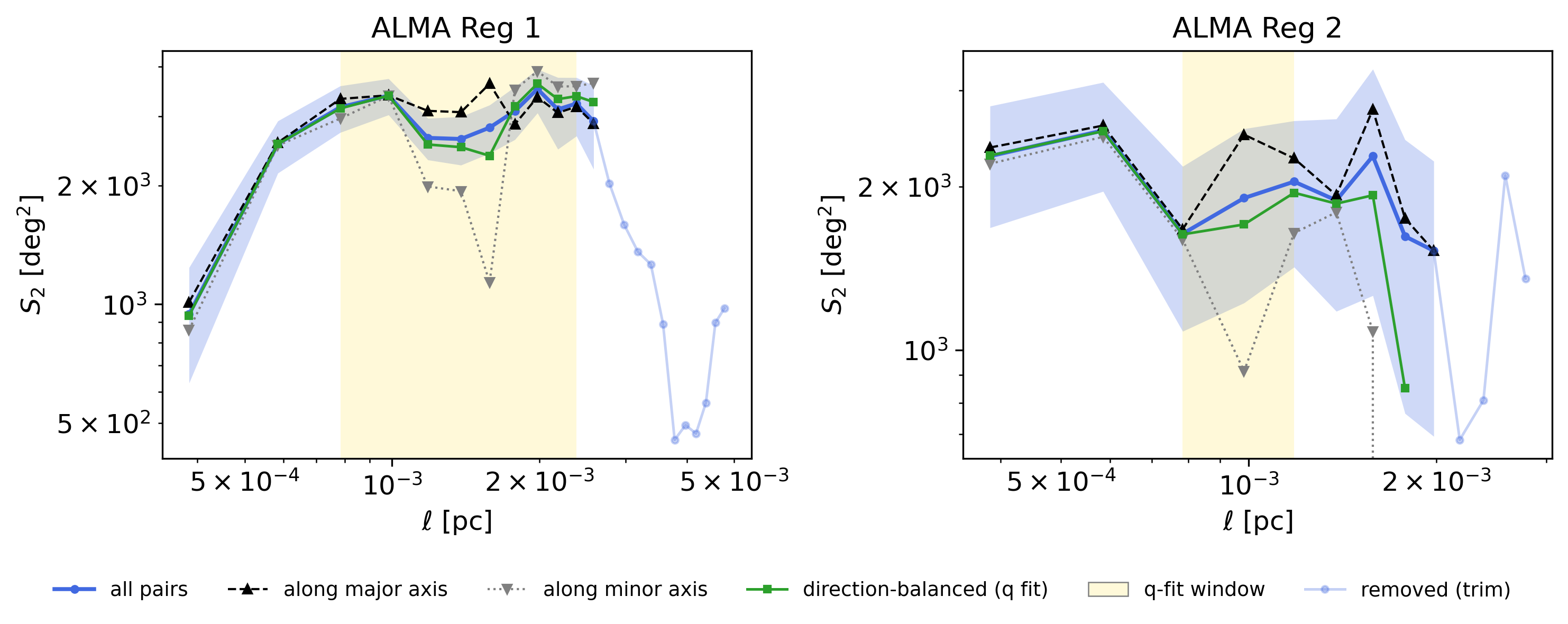}
    \caption{As Fig.~\ref{fig:S_2_functions_NGC1333}, but for
the two IRAS\,4A regions after re-binning. Faded points
show separations excluded by the trimming procedure,
including the large-separation downturn and all
subsequent bins.} 
    \label{fig:S_2_functions_IRAS4A}
\end{figure*}

\begin{figure*}
    \centering
    \includegraphics[width=0.48\linewidth]{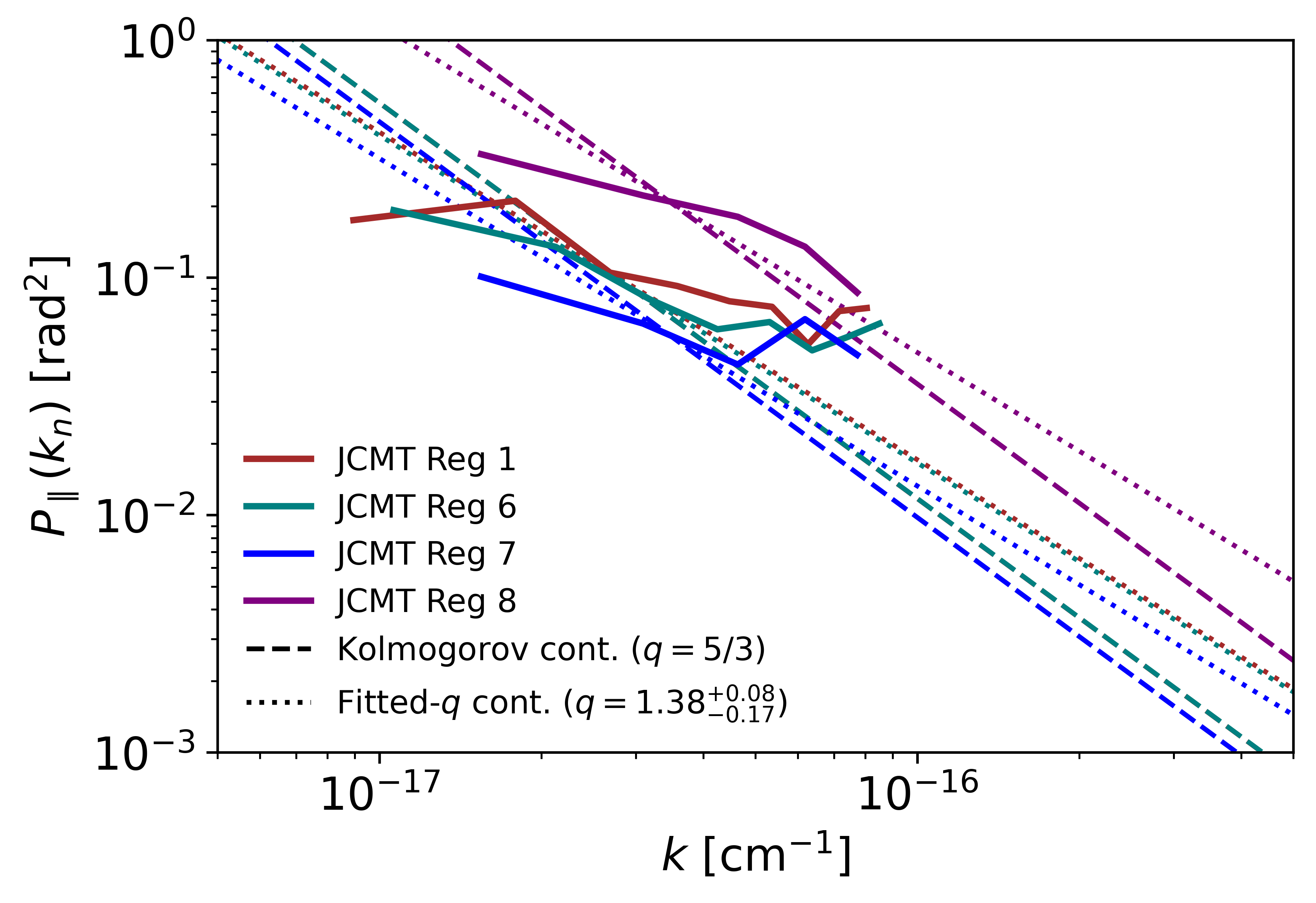}
    \includegraphics[width=0.48\linewidth]{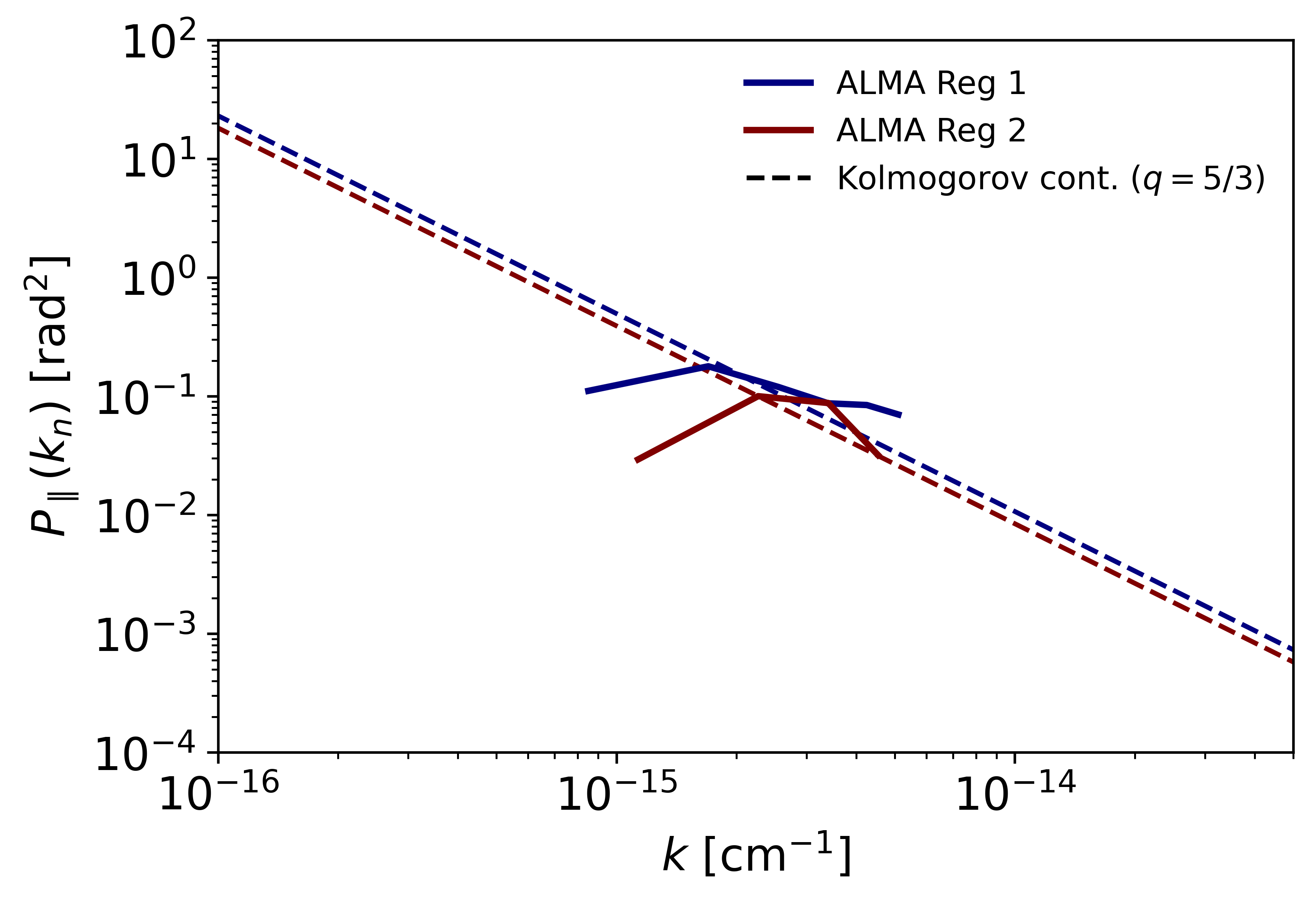}
    \caption{Fluctuation power per spectral bin,
$\mathcal{P}_\parallel(k_n)$ (Eq.~\ref{eq:Pk}), for the
NGC\,1333 regions (left) and IRAS\,4A regions (right).
Solid lines show the spectra derived from the retained
structure-function bins. Dashed lines show a Kolmogorov
spectrum ($q=5/3$). Dotted lines in the left panel show
the spectrum with the joint NGC\,1333 index,
$q=1.38^{+0.08}_{-0.17}$. Both reference spectra are
normalised to the measured spectrum at the geometric
mean of the retained $k_n$. These normalisations are
illustrative, since the continuations of $D(E)$ are anchored
separately at the lower edge of each data-informed interval.
}
    \label{fig:p_k_results}
\end{figure*}

We estimate the magnetic-field fluctuation power spectra, $\mathcal{P}_\parallel(k_n)$, from the modulus of
the Fourier transform of the binned structure function 
(Eq.~\ref{eq:Pk}). Every separation bin therefore contributes
to every mode. Contributions from separations that do not
represent the fluctuation statistics cannot be isolated by
discarding individual Fourier modes. We instead exclude these
separations from the pair sample before constructing the
structure function used in the transform. We apply a geometric
criterion to NGC\,1333 and a criterion based on the shape of
$S_2$ to IRAS\,4A. Figures~\ref{fig:S_2_functions_NGC1333}
and~\ref{fig:S_2_functions_IRAS4A} show the structure functions
and excluded separations. The resulting fluctuation spectra
are shown in Fig.~\ref{fig:p_k_results}.

For each NGC\,1333 region, we exclude pairs separated by more
than the full minor axis of its fitted ellipse, $2b$
(Table~\ref{tab:para1}). Beyond this separation, pairs cannot
span the region along its minor axis and preferentially
sample directions towards the major axis. The resulting
$S_2$ increasingly reflects structure along the elongated
region, including variations in the ordered field
(Fig.~\ref{fig:S_2_functions_NGC1333}). The largest separations
also contain few pairs, and including them introduces
oscillatory ringing in $\mathcal{P}_\parallel(k_n)$.
We find that this effect is most severe in Region~7, where the features change
position when the maximum separation is varied (not shown).
Such behaviour is consistent with ringing caused by the
finite analysis window, whose length sets the characteristic
spacing of the features. We apply the lag cap, retaining only pairs with separations
$\ell\leq2b$, before selecting the bin width and computing
the Fourier transform. Thus, only retained pairs determine
the bin width and contribute to the transformed structure
function. The magnetic-field strength $B$ is still estimated
from all vectors in each region. 

{In both IRAS~4A regions, $S_2$ falls below its saturation level at the
largest separations (see large $\ell$-value drop in $S_2$, in Fig.~\ref{fig:S_2_functions_IRAS4A}). These pairs mostly straddle
the dense core of the IRAS\,4A system, and are configured in a point-symmetric (hourglass) field such that pairs can have
similar orientations. The downturn in $S_2$ is therefore not guaranteed to trace the  fluctuation
statistics we intend to isolate in our analysis. We identify the start of the downturn as the first bin where 
$S_2$ lies more than $2\sigma$ below the maximum reached at smaller
separations, where $\sigma$ is the standard deviation of the difference 
between the two bins over the bootstrap realisations 
(Sec.~\ref{sec:diff_results}). That bin and all larger separations are
removed, including any later rises in $S_2$. The pairs are then cut at this 
separation and re-binned as described in Sec.~\ref{sec:struc_to_d_e}, so that
the excluded separations do not set the bin width, and the test is repeated
on the new bins. We further require that removing one additional bin changes
$D(E)$ by less than $2\sigma$ at every energy covered, and that more
than three modes remain in the analysis region. For Region~1, the downturn begins at $2.9\times10^{-3}$\,pc. Re-binning the
retained pairs reduces the bin width from 1.3 to 0.7 pixels, and the
stability test removes one further bin, leaving 12 bins (6 modes) out to 
$2.6\times10^{-3}$ pc. Without re-binning, removing the downturn would have 
left this region with only three modes. For Region 2, the downturn begins at 
$2.3\times10^{-3}$\,pc. The bin width is unchanged at 0.7 pixels, and 9 bins 
(4 modes) remain, out to $2.0\times10^{-3}$ pc.} 

For both datasets, we retain only regions with more than
three spectral modes after these selections. With fewer
modes, the spectrum is estimated from too few structure-function
bins to distinguish its shape reliably from the effects of
the finite sampled range. NGC\,1333 Regions~2, 3, 4, 5,
and~9 retain three or fewer modes and are therefore excluded
from the diffusion analysis. They are listed as ``not
measured'' in Table~\ref{tab:diffusion}. Both IRAS\,4A
regions satisfy the mode requirement.

We estimate the spectral index from
$S_2(\ell)\propto\ell^{q-1}$, corresponding to
$\mathcal{P}_\parallel(k)\propto k^{-q}$ for $1<q<3$.
We use separations of at least 2.5 beam FWHM.
Because the regions are elongated, the relative numbers
of pairs along and across the major axis change with
separation. We therefore fit a direction-balanced structure
function, which is the mean of the structure functions calculated
separately for these two directions. Only bins containing
at least 10 pairs in each direction are included.
We estimate the slope in logarithmic space using the
Theil--Sen method \citep{Theil1950PKNAW,Sen1968JASA},
which takes the median slope between all pairs of bins 
and reduces sensitivity to individual deviant bins.
Uncertainties are obtained by bootstrap resampling
(Appendix~\ref{app:uncertainty}).

The individual NGC\,1333 structure-function slopes are only
weakly constrained, differing from zero at approximately
1.5--1.9$\sigma$ (Table~\ref{tab:diffusion}). However, the
four retained regions have broadly similar fluctuation-spectrum
shapes, and their inferred indices are consistent with a
common value ($\chi^2=4.7$ for 3 degrees of freedom,
$p=0.20$). We therefore combine the regions to obtain a
better-constrained spectral index for their continuations
to unresolved scales. We pool the pairwise slopes calculated within each region
and take their median. Using only within-region slopes
allows each region to have a different structure-function
amplitude. The resulting index is
$q=1.38^{+0.08}_{-0.17}$, corresponding to a
structure-function slope that differs from zero at
3.4$\sigma$. Its bootstrap uncertainty bounds, 1.21--1.46,
lie below the Kolmogorov value. The remaining NGC\,1333
regions contain too few bins between 2.5 beam FWHM and
the lag cap to constrain a slope reliably. 

In IRAS\,4A, neither structure function rises significantly
beyond 2.5 beams, corresponding to approximately 160\,au.
The formal estimates are $q=1.08^{+0.21}_{-0.20}$ and
$1.43^{+1.11}_{-1.10}$ for Regions~1 and~2, respectively.
Both are consistent with a flat structure function
($q-1=0$), so we do not interpret them as measurements
of the fluctuation-spectrum index.

\subsection{Dependence of the diffusion coefficient on regional properties}
\label{app:D_vs_para}

\begin{figure*}
    \centering
    \includegraphics[width=\textwidth]{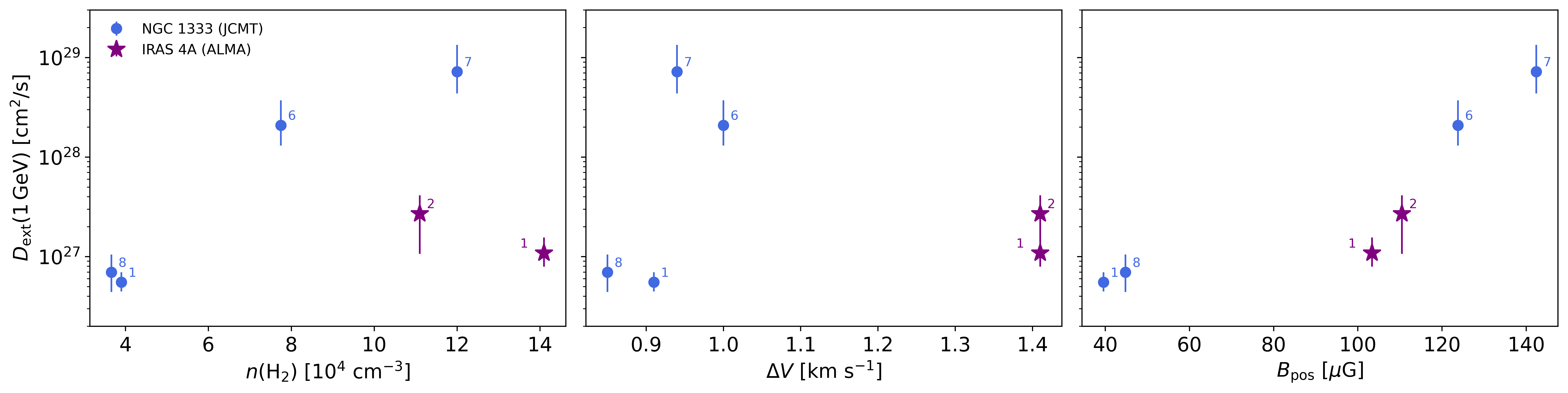}
    \caption{Kolmogorov-extrapolated diffusion coefficient at
1\,GeV against the H$_2$ volume density, non-thermal
linewidth, and plane-of-sky magnetic-field strength
(Table~\ref{tab:para1}) for the analysed regions.
Symbols, labels, and error bars are as in
Fig.~\ref{fig:ang_dis}.} 
    \label{fig:correlation_results}
\end{figure*}

Figure~\ref{fig:correlation_results} shows the
Kolmogorov-extrapolated diffusion coefficient,
$D_{\rm ext}(1\,{\rm GeV})$, against the H$_2$ volume density,
non-thermal linewidth, and plane-of-sky magnetic-field
strength, $B_{\rm pos}$. These quantities enter the scattering
calculation through the DCF field estimate and the Alfv\'en
speed. Within NGC\,1333, $D_{\rm ext}(1\,{\rm GeV})$
increases with all three, i.e. Regions~6 and~7 have higher
densities, linewidths, and field strengths, as well as
larger inferred diffusion coefficients, than Regions~1 and~8. 
These trends are coupled because the DCF estimate of $B$
depends on $n({\rm H_2})$, $\Delta V_{\rm nt}$, and
$\sigma_\phi$. For a common spectral shape and resolution,
the Kolmogorov continuation gives
$D_{\rm ext}(1\,{\rm GeV})\propto
n^{4/3}\Delta V_{\rm nt}^{5/3}\sigma_\phi^{-11/3}$
(Sec.~\ref{sec:dependence_on_cloud_properties}).
The NGC\,1333 regions with higher densities and linewidths
also have smaller angular dispersions. Their larger
coefficients therefore reflect the combined dependence on
these quantities. The steep trend with $B_{\rm pos}$ follows
from the same dependence, so we do not fit it separately. 

The IRAS\,4A regions lie below the NGC\,1333 trends.
Their field strengths are comparable to those of Regions~6
and~7, and their densities and line-widths are among the
highest in the sample. Nevertheless, their extrapolated
coefficients are approximately an order of magnitude or
more lower. Their larger angular dispersions and inferred
fluctuation power contribute to this difference, although
the two datasets also sample different spatial scales.
The comparisons therefore illustrate how the inferred
scattering depends on several regional properties together. 
They do not isolate an independent dependence on density,
linewidth, or field strength. 

\section{Uncertainty propagation}
\label{app:uncertainty}

{We estimate the uncertainties in our analysis by bootstrap resampling of the polarisation vectors in each  region~\citep{Efron1979AnSta}. Each of the $N_{\rm b}$ realisations draws as many vectors as the region contains, at random with replacement, so that some vectors appear more than once (and, on average, about a third are not drawn at all). All pairs are then
re-binned in the original separation bins, with a pair $(a, b)$ weighted by
$w_a w_b$, where $w_a$ is the number of times vector $a$ is drawn. A vector
drawn more than once is not paired with itself. Realisations that leave a
separation bin empty are redrawn. The bootstrap assumes that the vectors are independent. This holds approximately here, because the maps are binned to pixels close to the beam size (see Sec.~\ref{subsec:pola}).} 

{From each realisation we recompute $S_2(\ell_j)$ and
$\mathcal{P}_\parallel(k_n)$ (using Eq.~\ref{eq:Pk}), and re-derive the angular dispersion $\sigma_\phi$ and the magnetic field strength $B$ from the same resampled vectors. Each realisation is also
paired with a source distance drawn by Monte Carlo sampling from a normal distribution with a mean 
of 294 pc and standard deviation 17 pc \citep{Zucker2018ApJ}. The distance
rescales the mode wavenumbers, $k_n \propto 1/d$, but does not affect
$\mathcal{P}_\parallel(k_n)$ itself. 
  $D(E)$ is then computed for every
realisation. Because the bootstrap resamples the measured angles, which
already contain their measurement noise, no distribution needs to be assumed
for the angle uncertainties. These enter only through the noise-floor
subtraction in $\sigma_\phi$ and $S_2$ (see Sec.~\ref{sec:strength}
and~\ref{sec:struc_to_d_e}).} 

\begin{figure*}
    \centering
    \includegraphics[width=0.8\textwidth]{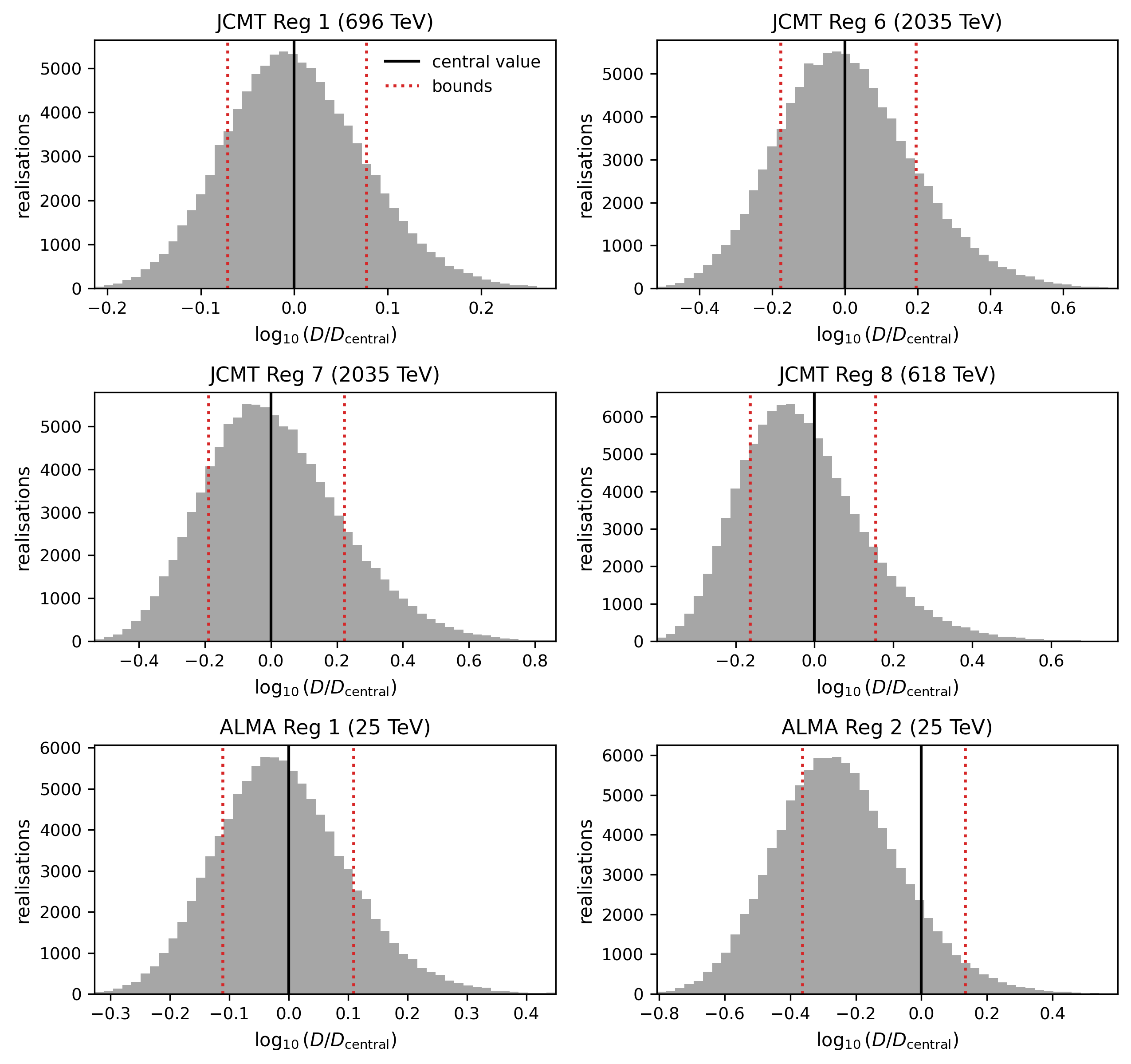}
    \caption{Bootstrap distributions of $D$ at the midpoint
of each region's data-informed interval, shown 
relative to the value calculated from the measured 
structure function (solid line). Dotted lines mark the adopted 
bounds, which are the 32nd percentile of the realisations below the central value and
the 68th percentile of those above it. For IRAS~4A Region~2, most
realisations lie below the central value.} 
    \label{fig:boot_dist}
\end{figure*}

{The central value of $D(E)$ is computed from the measured $S_2$ rather than from the bootstrap distribution. This is because the Fourier amplitudes of the noisier resampled structure functions are biased high for modes with little
power, which would
cause $D$ to be biased towards low values. This bias is largest where the measured spectrum has little power at the
resonant modes. It is most severe for IRAS~4A Region~2, whose structure function 
is nearly flat and approximately 85-95 per cent of its realisations give a lower $D$ than 
the measured value, while a minority extend to much higher values where
resampling happens to remove power from the relevant modes. Its bounds are 
therefore strongly asymmetric, and its $D$ is only weakly constrained (see Fig.~\ref{fig:boot_dist}). At each energy, the lower bound on $D(E)$ is the 32nd percentile of the
realisations that fall below the central value, and the upper bound is the
68th percentile of those above it, so that each bound encloses 68 per cent of
the realisations on its side of the central value. When the central value is
the median of the realisations, these are the 16th and 84th percentiles. This approach guarantees that our bounds always bracket the central value, and
that they reflect asymmetric uncertainties. These bounds are statistical, so they do not include the systematic uncertainties in $n({\rm H}_2)$, $\Delta V$ or the DCF method (see Appendix~\ref{app:limitations}). The continuations below each
data-informed band carry the fractional bounds of $D$ at the lower edge of
the band. For the fitted-$q$ continuation (see Fig.~\ref{fig:D_err_fitted_q}), the plotted uncertainty 
envelope also spans the uncertainty range of the joint index, defined in the same way.} 

{Uncertainties on the fitted $q$ index are obtained in the same way. The Theil-Sen slope (see Appendix~\ref{sec:modes}) 
is refitted to the direction-balanced $S_2$ of each realisation, in the same
bins, and the quoted uncertainty range is defined in the same way, around the measured value. For the joint fit to the NGC~1333 regions, the joint estimator 
(Sec.~\ref{sec:field_structure}) is applied to each realisation following the same method.}

{We use $N_{\rm b} = 10^5$ realisations for all regions, which is sufficient
for the bounds to converge. The two independent halves of the realisations
($5\times10^4$ each) give bounds that agree to within 1.5 per cent of $D$
for every region and energy. The exception is the upper bound for IRAS~4A
Region~2, which agrees to within 3 per cent because it lies in a sparsely
populated part of that region's distribution. The bounds on $q$ agree to
within $\pm0.02$. By comparison, with $10^3$ realisations the upper bounds still vary by up to
$\approx13$ per cent of $D$.} 

\section{Limitations}
\label{app:limitations}

The framework presented in this work is intended as a first demonstration of how CR diffusion coefficients can be constructed from observed magnetic-field structure traced by polarisation data. 
The main limitations arise from the finite spatial range sampled by the observations, the use of dust polarisation as a tracer of the magnetic field, and the methodological choices required to convert the measured field structure into a diffusion coefficient. 
We discuss each of these below.  

\textbf{Finite energy band and extrapolation:} our method recovers the magnetic fluctuation spectrum only over the finite interval of spatial scales sampled by the polarisation data. Through the resonant mapping between CR energy and magnetic fluctuation scale, this defines the energy band where $D(E)$ is directly informed by the data. 
Outside this range, the diffusion coefficient is a controlled extrapolation
based on the assumed unresolved spectrum, and the quoted GeV values depend
mainly on that assumption. For NGC~1333, the Kolmogorov and measured indices
give values that differ by $\sim1.5$ dex at 1 GeV
(Sec.~\ref{sec:D_regions}). 

\textbf{Local interpretation of the diffusion coefficient:} the quantity we construct and identify as a ``diffusion coefficient'' characterizes the local CR scattering environment implied by the measured magnetic fluctuations. Strictly, it can only be interpreted as a transport coefficient on scales larger than the corresponding scattering mean free path (see Sec.~\ref{sec:validity_energy_range}). On smaller scales, the particle motion would be quasi-ballistic rather than diffusive. Our results therefore capture spatial variations in the microphysics that sets CR scattering, but should not be interpreted as directly describing diffusive transport within the sampled region itself. 

\textbf{Magnetic focusing and mirroring:} the ordered field in IRAS~4A may
also modify the CR distribution through magnetic focusing and mirroring
\citep{Padovani2011A&A, Silsbee2018ApJ}. These effects enter the large-scale
transport problem separately from the local pitch-angle scattering
coefficient derived here. At the directly sampled TeV--PeV energies, energy
losses across the observed columns are negligible, and focusing and
mirroring approximately cancel for a smoothly varying field with a single
maximum \citep{Silsbee2018ApJ}. The polarisation data do not constrain the
three-dimensional field-strength profile needed to identify magnetic pockets
or calculate a mirror ratio, so we apply no geometrical correction to the
local scattering coefficient, and treat more complex field configurations as
a limitation of its cloud-scale interpretation.

\textbf{Quasi-linear and spectral representation of the turbulence:} our analysis converts the polarisation-angle structure function into a magnetic fluctuation power spectrum through a Fourier transform, and uses this as the input for a quasi-linear CR scattering estimate. 
The conversion involves two approximations. First, we use the modulus
of the transform (Eq.~\ref{eq:Pk}) rather than the cosine transform implied
by the relation between $S_2$ and $\mathcal{P}_\parallel$. This avoids
negative powers and removes the dependence on the offset of the first
separation bin. The two agree at the lowest modes, but the modulus 
overestimates the power at higher $k_n$, so $D$ is somewhat underestimated at
the low-energy end of each data-informed interval. Second, $S_2$ is not
sampled at separations below about one pixel, so fluctuation power on those
scales is not captured. Where the fluctuations decorrelate within a few
pixels, as in IRAS\,4A, this underestimates the power and hence overestimates
$D$.
    Our approach treats the fluctuations as a superposition of independent linear modes. Interstellar turbulence is intrinsically non-linear and intermittent~\citep{Hennebelle2012A&ARv, Falceta2014NPGeo}, so this representation does not retain phase correlations, coherent structures, or non-stationary features that may also influence CR scattering. 
    {In particular, quasi-linear theory requires weak, incoherent fluctuations that
randomise the CR gyrophase. Strong ($\delta B/B \gtrsim 1$) or coherent
structures, such as shocks, current sheets or non-resonant instabilities, can
sustain gyrophase correlations and lead to non-gyrotropic, non-diffusive
transport \citep{Bell2004MNRAS, Shalchi2009ASSL}. Our angular dispersions, $\sigma_\phi \approx 0.4$--$0.8$ rad for the
analysed regions, correspond to $\delta B/B$ of the same order, which places
some regions close to this limit. However, these values are upper limits on the
fluctuation amplitude, since $\sigma_\phi$ also includes large-scale curvature
of the field (see magnetic-field strength estimates below).}  
    Methods that relax this assumption, such as the Hilbert-Huang transform and related empirical-mode techniques~\citep{Huang1998RSPSA, Huang2008RvGeo}, may provide a more faithful decomposition for future datasets. However, at present, the finite dynamic range and signal-to-noise of the 
    current polarisation maps 
    do not yet justify a more complex non-linear decomposition, so the Fourier-based treatment is an appropriate first-order choice. 
    
\textbf{Projection, depth, and beam-averaging effects:} dust polarisation is 2-dimensional, and only traces the plane-of-sky component of the magnetic-field orientation, integrated through the emitting material. 
    Line-of-sight averaging can suppress the apparent angular dispersion, while any anisotropic structure along the line of sight is not directly recovered. The line-of-sight field can be constrained independently. Faraday rotation of background radio sources, combined with extinction maps and chemical models of the electron abundance, gives the direction and strength of the line-of-sight field across molecular clouds. Together with dust polarisation, this has been used to reconstruct the 3D field morphology of the Perseus cloud on pc scales \citep{Tahani2018A&A, Tahani2022A&A}. 3D dust maps can now also resolve the gas distribution along the line of sight up to pc resolution
\citep{Leike2020A&A, Zucker2021ApJ, Edenhofer2024A&A}, and could replace the fixed depth we assume for gas densities on these scales. Because Faraday rotation is weighted by the free-electron
density, it mainly traces cloud envelopes, so Zeeman measurements
\citep{Crutcher2012ARA&A} remain needed for the line-of-sight field in the
dense gas analysed here. Combined with sub-pc polarisation data, these would allow the scattering environment to be characterized in three dimensions.

    In optically thick regions, the polarized emission may also be weighted toward the $\tau \sim 1$ surface rather than the full scattering volume, with the effective depth varying systematically with density.  While the trends of $D$ with density within NGC\,1333 follow from the DCF and scattering dependencies (see Appendix~\ref{app:D_vs_para}), our results should be interpreted with this filtering in mind, which could imprint a density-correlated observational bias, or mask any underlying physical dependencies that emerge in regions where $\tau \gg 1$. 
    Beam averaging further smooths sub-beam angular structure, reducing the measured fluctuation power on the smallest resolved scales, which are most relevant to the low-energy end of the directly probed $D(E)$ band (see Sec.~\ref{sec:d_e_physical_features} and Eq.~\ref{eq:e_band_low}). {In our analysis, the finest modes correspond to wavelengths of about 1.2 beam
FWHM for NGC\,1333 and 1.3 for IRAS\,4A, where this smoothing suppresses the
measured power, and the continuations below the data-informed intervals are
anchored at this edge.} 
    These effects generally act to underestimate the true magnetic fluctuation power and, hence, could underestimate the CR scattering rate. The resulting diffusion coefficients we obtain should therefore be regarded as effective plane-of-sky upper-estimates. 
    
\textbf{Magnetic-field strength estimates:} the magnetic-field strength enters our calculation through the DCF method (see Sec.~\ref{sec:strength}; also~\citealt{Davis1951, CFmethod1953}), which carries well-documented uncertainties. We adopt standard correction factors calibrated against MHD simulations~\citep[e.g.][]{Liu2021ApJ}, but the underlying assumption of approximate equipartition between turbulent kinetic and perturbed magnetic energy densities is statistical rather than strictly local~\citep{Chen2022MNRAS}.  It may break down in dense, self-gravitating cores below 0.1 pc scales, where the gas flows are no longer governed by pure MHD turbulence~\citep{Liu2021ApJ}. Alternative estimators that retain compressible modes can differ from DCF by factors of a few~\citep{Skalidis2021A&A, Liu2022ApJ}. 
{Moreover, density, linewidth, and polarisation measurements do not
necessarily weight exactly the same material along the line of sight. This
applies particularly to IRAS\,4A, whose linewidth is taken from NH$_3$
observations with a 32$''$ beam, much larger than the ALMA regions and
including in-fall and outflow motions, and whose density assumes the same
0.1 pc depth as the NGC~1333 regions, much larger than the cores. Adopting
instead a depth comparable to the core size and a core-scale linewidth, as
in \citet{Girart2006Sci}, would increase $B$ for IRAS\,4A by a factor of about
two.} These uncertainties affect the normalization of $D(E)$, and the mapping between CR energy and resonant magnetic-field scale. {At fixed density and fluctuation power, $D$ scales approximately as $B^2$,
so a factor-of-two uncertainty in $B$ corresponds to a factor of about four
in $D$, while the energies of the data-informed intervals scale linearly
with $B$ (Eqs.~\ref{eq:e_band_low} and~\ref{eq:e_band_high}). The ionization fraction affects only the
normalization as $D \propto \xi^{1/2}$ through the Alfv\'en speed of the ionized gas. This means that a value at the
lower end of the range found in dense cores, $\xi = 10^{-7.5}$ \citep{Caselli1998ApJ}, would lower $D$
by a factor of $\sim 3$. The angular dispersion also includes any large-scale curvature of the ordered
field within a region. Measuring the dispersion about a smooth
(second-order polynomial) model of the ordered field instead lowers
$\sigma_\phi$ by 24-50 per cent for the analysed NGC\,1333 regions and by
14 per cent for IRAS\,4A Region 1, and leaves Region~2 unchanged. We do not
remove this contribution, because the curvature cannot be cleanly separated
from turbulent fluctuations on the largest scales, but $B$ may be
correspondingly underestimated, by up to a factor of about two in our most
curved region, NGC~1333 Region 7. Separating the two reliably would require
a dedicated model of the ordered field.} 

\textbf{Polarisation efficiency and dust grain alignment:} our analysis assumes that polarized dust emission reliably traces the local magnetic-field orientation. This requires efficient grain alignment. The favoured mechanism is the radiative alignment of torques~\citep[][]{Lazarian2007MNRAS, Andersson2015ARA&A}. In shielded dense gas, attenuation of the radiation field can reduce the grain alignment efficiency, which may account for the observed decrease of polarisation fraction with column density in dense cores~\citep[e.g.][]{Whittet2008ApJ, Alves2014A&A}, although alignment may persist to relatively high densities in some environments~\citep{Pattle2019ApJ}. This introduces a density-dependent loss of magnetic-field information, weighting the recovered structure toward lower-density and/or more strongly illuminated material. In our analysis, this may act to bias the measurement of $D(E)$ away from the highest-density gas along a line-of-sight. 

\section{Comparison with complementary CR constraints}
\label{app:comparison}

\subsection{Ionization-rate constraints in NGC~1333}
\label{app:low_energy_measurements}

Complementary constraints on lower-energy CRs in NGC~1333 are provided by
measurements of the CR ionization rate. \citet{Luo2026arXiv} used H~I narrow
self-absorption (HINSA) to infer a CR ionization rate of
$\zeta \simeq 0.4$--$2\times10^{-17}$\,s$^{-1}$, with a decrease toward
increasing molecular-hydrogen column density. At a representative column
density of $N({\rm H}_2)\sim10^{22}$\,cm$^{-2}$, however, the ionization
response is dominated by protons near 10~MeV. These measurements therefore
probe a much lower-energy CR population than either the \textit{Fermi}-LAT
measurements or the TeV-PeV energies directly sampled by our analysis.

\citet{Pineda2024A&A} mapped the CR ionization rate across part of NGC~1333
using the DCO$^+$/H$^{13}$CO$^+$ abundance ratio and the analytic relations
of \citet{Caselli1998ApJ}. They found typical values of
$\zeta \approx 10^{-16.5}$\,s$^{-1}$, somewhat higher than
the HINSA estimates. They also found enhancements around
embedded YSOs and a strong increase toward the northwest,
which they attributed to locally accelerated CRs or to
interaction with an expanding shell. 

The analytic relations assume steady-state ion chemistry, which may not be reached even in cold dense cores
\citep[e.g.][]{Pagani2013A&A}. In particular, the slowly
evolving ortho-to-para ratio of H$_2$ can remain far from
its thermal-equilibrium value and affect deuterium
fractionation in the centers of high-density cores \citep{Lin2025ApJ}. 
Their application also depends on temperature through
deuterium fractionation. In warm or recently
shocked gas, where deuteration through H$_2$D$^+$ is suppressed above
$\approx20$--30~K \citep{Caselli2012A&ARv}, the DCO$^+$/HCO$^+$ ratio can
decrease independently of the ionization rate, which the analytic relations
would interpret as a larger electron fraction and ionization rate. The local
enhancements occur where these assumptions are least secure, near
protostars, outflows and the shell. In these regions, shocks and heating can
reset the chemistry on timescales of $10^4$--$10^5$~yr. The adopted uniform excitation temperature and densities
interpolated from $18''$ \textit{Herschel} maps may also 
represent these regions poorly. These effects complicate the interpretation of the spatial
variations in the inferred ionization rate. We therefore
do not use those variations to constrain CR transport
in this work. 

\subsection{CR transport in other molecular clouds}
\label{app:other_clouds}

$\gamma$-ray observations of passive giant molecular clouds find CR spectra
broadly consistent with a uniform Galactic CR sea 
\citep{Aharonian2020PhRvD,Peron2022A&A}. Departures in individual systems
can arise either from an enhanced or depleted incident CR population, or
from energy-dependent filtering during transport through the cloud.
Supernova-remnant--cloud systems provide evidence for 
enhanced CR irradiation. The pion-decay
feature of IC~443 and W44 reveals protons accelerated or re-accelerated near
dense gas \citep{Ackermann2013Sci}, extended emission around W28 traces
escaped CRs \citep{Aharonian2008A&A,Gabici2009MNRAS}, and enhanced
ionization rates and C$^0$/CO ratios near W51C and W28 show that CRs of
${\sim}100$~MeV also reach the surrounding gas
\citep{Ceccarelli2011ApJ,Vaupre2014A&A,Yamagishi2023PASJ}.

The penetration of these CRs depends on the magnetic
conditions within and around the cloud. Shock-clump interactions, as in RX~J1713.7$-$3946, generate a
turbulent, magnetically amplified layer around dense clumps 
\citep{Inoue2012ApJ,Sano2013ApJ,Sano2020ApJ}, and the resulting
energy-dependent penetration modifies the $\gamma$-ray spectrum produced
inside them \citep{Gabici2014MNRAS,Celli2019MNRAS}. For the gyro-resonant scattering considered in our framework,
the scattering efficiency depends on the fluctuation power
at the relevant resonant scales. CR streaming can generate such fluctuations
through the Bell instability, hampering the penetration of particles below
${\sim}1$~TeV \citep{Inoue2019ApJ}, so that low-energy CR ionization and
Fe~K$\alpha$ emission may be concentrated near cloud boundaries
\citep{Fujita2021ApJ}. Pre-existing slow-diffusion layers can similarly
produce substantial attenuation \citep{Ng2026PASJ}. 
In such systems, $\gamma$-ray and ionization measurements constrain the CR
population that reaches the gas, whereas our method constrains the magnetic
fluctuations that regulate that access.

\subsection{CR transport in other high-energy environments}
\label{app:other_HE_environments}

Extended TeV halos around middle-aged pulsars imply diffusion coefficients
strongly suppressed relative to the Galactic average, at energies close to
those sampled here \citep{Abeysekara2017Sci}. These systems trace CR leptons
escaping powerful accelerators, however, which may themselves generate the
turbulence that confines them \citep{Evoli2018PhRvD}, and are therefore not
directly comparable with scattering by fluctuations already present in
molecular clouds.

Young stellar clusters provide closer comparisons. Extended $\gamma$-ray
emission around the Cygnus Cocoon and Cyg~OB2, Westerlund~1 and~2, and
younger embedded regions such as NGC~3603, W40, RCW~38 and clusters in the
Vela molecular ridge
\citep{Aharonian2019NatAs,Abeysekara2021NatAs,Aharonian2022A&A,
Peron2024NatAs,Pandey2024ApJ} 
provides constraints on the radial CR-density profiles. 
These include inferred ${\sim}1/r$ declines around 
Cyg~OB2 and Westerlund~1, consistent with continuous injection and
approximately homogeneous diffusion over pc to tens-of-pc scales,
at proton energies approaching the PeV range. Such smooth large-scale
profiles do not require a spatially uniform local scattering rate.

The Galactic Centre provides a further illustration of this. Diffuse multi-TeV
emission from the Central Molecular Zone follows a ${\sim}1/r$ CR profile
without significant spectral variation, interpreted as quasi-steady
injection followed by approximately homogeneous diffusion over tens to
hundreds of pc \citep{HESS2016Natur,Devin2025arXiv}. Yet MeerKAT
resolves numerous narrow non-thermal filaments, revealing a highly
structured magnetic environment on much smaller scales
\citep{Heywood2022ApJ,YusefZadeh2022ApJ}, while the filament morphologies
depend strongly on the local transport parameters
\citep{RichterAddo2026arXiv}.

These observations allow an approximately homogeneous
description of transport on large scales despite substantial
magnetic structure on smaller scales. Our results reveal
local variations in scattering that such an effective
description could conceal. How these variations combine
to determine the large-scale coefficient depends on the
paths CRs take through the medium, with the most strongly
scattering regions potentially exerting disproportionate
control \citep{Ewart2026MNRAS}. 

\end{document}